\documentclass[twocolumn,showpacs,preprintnumbers,amsmath,amssymb]{revtex4}

\usepackage{graphicx}
\usepackage{dcolumn}
\usepackage{bm}

\begin{document}

\preprint{APS/}

\title{High-temperature superconductivity and normal state in the Holstein-t-J model}

\author{E.E.  Zubov}
\affiliation{%
O.O. Galkin Institute for Physics and Engineering NASU\\
72, R. Luxemburg Str.,Donetsk, 83114, Ukraine}%
\date{\today}

\begin{abstract}
A possible origin  of the high-temprature superconductivity in cuprates has been suggested. It is supposed that  electron-phonon interaction determines the strong  correlation narrowing of the  electron band. It provides the conditions for the formation of a singlet electron pair coupled by exchange interaction. For the pure t-J model it has been proved that these electron pairs are destroyed by a strong effective kinematic field. The detailed analysis of  an influence of  the Holstein polaron excitations  upon normal and superconducting properties  of the strongly correlated electrons was made. A calculated critical temperature of the superconductivity and gap function are in good agreement with experimental data for cuprates.  
\end{abstract}

\maketitle

\section{Introduction}

In the 21st century a phenomenon of the high-temperature superconductivity in cuprates continues to attract the attention of many researches \cite{Shin}. There are a tremendous number of suggested mechanisms of this phenomenon. In any case there is not  a such theory  which  would describe  all properties of  this complicated  state. In this work we have centered on the main pecuilarity which in our opinion might help to illuminate the origin of high -T$_{\textit{SC}}$  in cuprates. The electron-phonon coupling is supposed to be not essential in the Cooper electron pairings. But this interaction forms the polaron excitations which play an imortant role in the correlation narrowing of the  electron band. In that case it is necessary to differentiate the collectivized electrons in metals  and ones  in doped cuprates. Indeed, in metals there is wave electron states with a possibility of the  site double occupancy. But their hole states is virtual. And that's why  we have the partition function $\exp (\varepsilon _{\textbf{k}\sigma } /T) + \exp (\varepsilon _{\textbf{k} - \sigma } /T)$ for electron excitations $\varepsilon _{\textbf{k} \pm \sigma }$.  In cuprates a coordinate representation is realized for electron wave functions and we have the partition function $1 + \exp (\varepsilon _\sigma  /T) + \exp (\varepsilon _{ - \sigma } /T)$ with electron levels $\varepsilon _{ \pm \sigma } $ and hole state. 

The cuprates belong to class of the  strongly correlated electron system.  In work \cite{Gros01} an effective Hamiltonian   of the t-J model was suggested based on the use of Gutzwiller projection operator. It allowed to exclude the upper Hubbard band with double site occupancy by electrons and essentially to simplify an investigation of the  strongly correlated electron systems. In work \cite{Baskar}  a  mean field approximation  of the  t-J model was developed to study the high-temperature superconductivity. In this work a  fundamental idea  about spin pairing via electron exchange interaction  was formulated. Unfortunately, authors  were not taken into account the essential difference between metal and strongly correlated electrons. Using Bogolyubov's u-v transform of the Hamiltonian they obtained the equation for gap function to be similar in BCS theory.
 
In this work we propose to divide the mean field  BCS type   Hamiltonian  into uniform and nonuniform parts. The perturbation theory was built with uniform unperturbed Hamiltonian. The nonuniform part  is  neglected  since it has a weak influence on the hopping integral.   A hopping term of the total t-J Hamiltonian is considered as perturbation in the limit of a weak doping  with $\\$ u-v transformed creation and destruction operators. The anormal mean values to be proportional the superconductive gap function were calculated. It has been obtained the condition on values of the chemical potential and exchange parameter. With account of the correlation band  narrowing  we make the conclusion about impossibility of  HTSC in the pure t-J model.

In what follow we include into consideration  the electron-phonon interaction. The evidences for a presence of one and its important role in the strongly correlated systems were emphasised  in works \cite{Kulic,Ciuchi,Alexandr}. In view of the fact that Hamiltonian of electron-phonon coupling is nonuniform many authors simplify the kinematic part  by simple renormalization of the hopping integral \cite{Alexandr} or use the theory of Eliashberg for collectivized metal electrons \cite{Kulic}. In former case it gives rise to drastic supression of the electron band and is responsible for the absence of  HTSC in a  system without interaction of polarons. The simplest form of the Holstein Hamiltonian for polarons  needs to be considered  with uniform electron-phonon interaction  and Einstein phonon mode. One can provide the exact unitary transform to separate fermion and boson degree of  freedom. It allows to build the subsequent perturbation theory  of the strongly correlated electron system with Holstein's polarons.
 
The structure of the paper is as follows. In Sec.II we consider the Hamiltonian of pure  t-J model in the superconducting state. It was separated the uniform mean field part with correponding coefficients of u-v transform. It enables us  in Sec.III to build the perturbation theory for strongly correlated electrons  in  the superconducting state. In particular, it was obtained the transformed  Hubbard operators in coordinate representation using the Bogolyubov's u-v transform. As a result, the equation for gap function and conditions for superconducting state were presented. In Sec. IV  the properties of normal state without electron-phonon coupling are considered. In the framework of the developed diagrammatic method it was shown the absence of  superconductivity in a  pure t-J model. In Sec.V. a normal state of the  cuprate d-electrons with polaron excitations is investigated to find the critical temperature of superconducting  state. In this section it has been solved   the problem of  a  frequency summation with infinity number of poles as implicit functions. The suggested method of an inverse function allowed to calculate the diagrammatic contributions for all polaron bands. In Sec.V the obtained equations are solved numerically that allowed to find the concentration dependences of the critical temperature  \textit{T}$_{\textit{SC}}$  and gap function $\Delta$   versus temperature. The theoretical values of \textit{T}$_{\textit{SC}}$ and  $\Delta$   are  in good agreement with experiment that supports the model to put forward by us.

\section{HAMILTONIAN OF THE SYSTEM}
The Hamiltonian of the Holstein model with strongly correlated electrons takes the form : 
\begin{eqnarray}
\hat H = \hat H_f  + \hat H_b
\,,
\label{energy01}
\end{eqnarray}
where the Fermi part, $\hat H_f$ , is expressed as follows :
\begin{eqnarray}
\hat H_f  = \sum\limits_{i,j} {J_{ij} \left( {{\bf S}_i {\bf S}_j  - \frac{1}{4}n_i n_j } \right) - \mu \sum\limits_{i\sigma } {n_{i\sigma } } }  + \hat V
\label{energy02}
\end{eqnarray}
Here,  $J_{ij}$ is the indirect exchange of the collectivized d-electrons with spins ${\bf S}_i$ and ${\bf S}_j$, $n_i  = n_{i\sigma }  + n_{i - \sigma }$ is the electron concentration on $\textit{i}$-site,  $\mu$  is the chemical potential. The perturbation $\hat V$  is written as
\begin{eqnarray}
\hat V = \sum\limits_{i,j,\sigma } {t_{ij} c_{\sigma i}^ +  c_{\sigma j} } (1 - n_{i - \sigma } )(1 - n_{j - \sigma } )
\,,
\label{energy03}
\end{eqnarray}
where $c_{\sigma i}^ +$($c_{\sigma j}$) creates (annihilates) an electron of spin $\sigma$  on lattice site $\textit{i}$ and  $t_{ij}$ is the hopping integral to be equal to $\textit{t}$ for nearest neighbours. The Hamiltonian ~(\ref{energy02})  of $\ $ t-J model  reflects the strong electron correlations. In a weak doping level we will  consider the part ~(\ref{energy03}) as a perturbation.

The boson part of  Hamiltonian ~(\ref{energy01}) has a form similar to that  used in the Holstein model  of a small polaron:
\begin{eqnarray}
\hat H_b  =  - g\sum\limits_i {n_i \left( {b_i^ +   + b_i } \right) + \omega _0 \sum\limits_i {b_i^ +  b_i } }
\,,
\label{energy04}
\end{eqnarray}
where $\textit{g}$ is the electron-phonon coupling strength,  $b_{i}^+$ and  $b_{i}$ are the phonon creation and destruction operators. We will use the  Einstein model where the phonon frequency  $\omega_0$ is assumed to be dispersion-free.

The Lang-Firsov unitary transform\cite{Lang} $\tilde U = \exp \left( {\tilde S} \right)$
  of Hamiltonian ~(\ref{energy04}) allows to separate the boson and fermion operators in ~(\ref{energy04}), where $\tilde S =  - \frac{g}{{\omega _0 }}\sum\limits_i {n_i (b_i^ +   - b_i )} $. As a result we have
\begin{eqnarray}
\hat{\tilde{H_{b}}}  = \tilde U^{ - 1} \hat H_{b} \tilde U = \omega _0 \sum\limits_i {b_i^ +  b_i }  - \xi \sum\limits_i {n_i }
\,,
\label{energy05}
\end{eqnarray}
where $\xi  = g^2 /\omega _0 $ is the polaron binding energy. The unitary transformed perturbation $\hat V$ is presented as 
\begin{eqnarray}
\hat V = \sum\limits_{ < ij > ,\sigma } {t_{ij} \tilde c_{i\sigma }^ +  \tilde c_{j\sigma } } (1 - n_{i - \sigma } )(1 - n_{j - \sigma } )
\label{energy06}
\end{eqnarray}
Here, the unitary transformated   Fermi operators
\begin{eqnarray}
\tilde c_{i\sigma }  = Y_i c_{i\sigma } 
\label{energy07}
\end{eqnarray}
are product of Bose  $Y_i  = e^{\lambda (b_i ^ +   - b_i )} $ and corresponding Fermi destruction operators,  where  $\lambda=g/\omega_0$.  It is necessary to point out that first and second terms of the Hamiltonian ~(\ref{energy02}) are not changed  under  transform  $\tilde U$.

One can separate in a Heisenberg part  of  the Hamiltonian ~(\ref{energy02}) by standard manner a mean field to be connected with anormalous averages \cite{Baskar}. Then an unperturbed Hamiltonian takes the form
\begin{eqnarray}
\hat H_{0f}  = \sum\limits_{ < ij > \sigma } {\left\{ {\Delta _{ij\sigma } c_{i\sigma }^ +  c_{j - \sigma }^ +   + \Delta _{ij\sigma }^* c_{i - \sigma } c_{j\sigma } } \right\}}  - 
\nonumber \\
-\sum\limits_{i\sigma } {\tilde \mu _\sigma  n_{i\sigma } }
\,,
\label{energy08}
\end{eqnarray}
where $\tilde \mu _\sigma   = \tilde \mu  - \sigma J(0) < S^z  > $ , $\tilde \mu  = \mu  + \xi $,  $ < S^z  > $ is a mean electron spin and   $\sigma= \pm 1$. The gap functions are expressed via exchange parameters:
\begin{eqnarray}
\begin{array}{l}
 \Delta _{ij\sigma }  =  - J_{ij}  < c_{i - \sigma } c_{j\sigma }  >  \\ 
 \Delta _{ij\sigma }^*  =  - J_{ij}  < c_{i\sigma }^ +  c_{j - \sigma }^ +   >  \\ 
 \end{array}
\label{energy09}
\end{eqnarray}

In a wave space the Hamiltonian ~(\ref{energy08}) takes the form
\begin{eqnarray}
\hat H_{0f}  = \sum\limits_{ \bm k \sigma } {\left\{ {\Delta _{\bm k\sigma } c_{\bm k\sigma }^ +  c_{- \bm k - \sigma }^ +   + \Delta _{\bm k\sigma }^* c_{- \bm k - \sigma } c_{\bm k\sigma } } \right\}}  - 
\nonumber \\
-\sum\limits_{\bm k \sigma } {\tilde \mu _\sigma  n_{\bm k\sigma } }
\,,
\label{energy10}
\end{eqnarray}
where the gap functions $\Delta _{\bm k\sigma }$ can be presented as
\begin{eqnarray}
\Delta _{\bm k\sigma }  = - \frac{1}{N}\sum\limits_{\bm q} {J(\bm q + \bm k)}  < c_{- \bm q  - \sigma } c_{\bm q \sigma } > 
\label{energy11}
\end{eqnarray}
and  $\Delta _{\bm k\sigma }^*$  is conjugate function $\Delta _{\bm k\sigma }$ . One can point out that in Eqs. ~(\ref{energy09}) and ~(\ref{energy11}) the  operators of creation and destruction are not transformed by operator $\textit{Y}_i$ from ~(\ref{energy07}). The Bogolyubov's  u-v transform
\begin{eqnarray}
\begin{array}{c}
\textit{c}_{\bm k\sigma }  = u_{\bm k\sigma }^* \alpha _{\bm k\sigma }  + v_{\bm k\sigma } \alpha _{ - \bm k - \sigma }^ +   \\ 
\textit{c}_{\bm k\sigma }^ +   = u_{\bm k\sigma } \alpha _{\bm k\sigma }^ +   + v_{\bm k\sigma }^* \alpha _{ - \bm k - \sigma } \\
\end{array}
\label{energy12} 
\end{eqnarray}
to new operators  $\alpha_{\bm k\sigma}$ and $\alpha_{-\bm k-\sigma}^+$   allows to diagonalize $\hat H_{0f}$  with the next conditions 
\begin{eqnarray}
u_{\bm k\sigma } = u_{ - \bm k - \sigma },\quad v_{\bm k\sigma } = - v_{ - \bm k - \sigma } , 
\nonumber \\
\quad \left| {u_{\bm k\sigma } } \right|^2  + \left| {v_{\bm k\sigma } } \right|^2  = 1
\label{energy13} 
\end{eqnarray}
Then we have
\begin{eqnarray}
\hat H_{0f}  = \sum\limits_{\bm k\sigma } {\tilde E_{\bm k\sigma } } \alpha _{\bm k\sigma }^ +  \alpha _{\bm k\sigma },
\label{energy14} 
\end{eqnarray}
where 
\begin{eqnarray}
\tilde E_{\bm k\sigma }  =  - \tilde \mu \sqrt {1 + \left( \frac{\Delta _{\bm k }}{\tilde \mu } \right)^2 } 
\label{energy15} 
\end{eqnarray}
In what follows we will consider a paramagnetic state when $ < S^z  >  = 0$. Then one can put 
\begin{eqnarray}
\tilde \mu _\sigma   = \tilde \mu ,\quad \left| {\Delta _{k\sigma } } \right| = \left| {\Delta _{ - k - \sigma } } \right| = \Delta _k 
\label{energy16} 
\end{eqnarray}

So far it has been obtained that the BCS Hamiltonian  ~(\ref{energy14}) coincides with similar Hamiltonian of Baskaran-Zou-Anderson \cite{Baskar}. Unfortunately, the authors of work \cite{Baskar} do not separate perturbation $\hat V$ from ~(\ref{energy01}). Instead of this they  narrow band multiplying  the hopping integral \textit{t} by factor \textit{x} to be equal  to  hole concentration.  It does not allow to find the rigorous statement relatively an appearance of the superconductivity since the band energy at \textit{x}$\approx 0$ has finite quantity.  That's why we will expand Eq.~(\ref{energy14}) in terms of the small parameter   up to third order:
\begin{eqnarray}
\tilde E_{\bm k}  =  - \tilde \mu \left\{ {1 + \frac{1}{2}\frac{{\Delta _{\bm k}^2 }}{{\tilde \mu ^2 }} - \frac{1}{8}\frac{{\Delta _{\bm k}^4 }}{{\tilde \mu ^4 }} + \frac{3}{{48}}\frac{{\Delta _{\bm k}^6 }}{{\tilde \mu ^6 }} - ...} \right\}
\label{energy17} 
\end{eqnarray}

Apparently, the corrections to chemical potential in Eq. ~(\ref{energy17})  will produce the additional nonuniform part   $\Delta H$  to perturbation  $\hat V$ from ~(\ref{energy03}) in the coordinate space:
\begin{eqnarray}
\Delta H = \sum\limits_{ij} {\delta t_{ij} \alpha _{i\sigma }^ +  \alpha _{j\sigma } },
\label{energy18} 
\end{eqnarray}
where
\begin{eqnarray}
\delta t_{ij} \approx \frac{1}{N}\sum\limits_{\bm k}{\left\{ { - \frac{1}{2}\frac{{\Delta _{\bm k}^2 }}{{\tilde \mu ^2 }} + \frac{1}{8}\frac{{\Delta _{\bm k}^4 }}{{\tilde \mu ^4 }} - \frac{3}{{48}}\frac{{\Delta _{\bm k}^6 }}{{\tilde \mu ^6 }}} \right\}e^{i{\bm k}(\bm {R}_i  - \bm {R}_j )}}
\nonumber \\
\label{energy19} 
\end{eqnarray}

In what follows our consideration is limited by the square lattice and s- or d- symmetry of order parameter $\Delta _{\bm k}  = \Delta (\cos k_x a \pm \cos k_y a)$. In ~(\ref{energy17}) we will extract one site part of $\Delta H_1 $:

\begin{eqnarray}
\Delta H_1  = \sum\limits_{ij} {\delta t_{ii} \alpha _{i\sigma }^ +  \alpha _{i\sigma } } 
\label{energy20} 
\end{eqnarray}
The rest  $\Delta H_2 $  of ~(\ref{energy18}) may be presented as
\begin{eqnarray}
\sum\limits_{i \ne j} {\delta t_{ij}  < \alpha _{i\sigma }^ +  \alpha _{j\sigma }  >  + } \sum\limits_{i \ne j} {\delta t_{ij} \alpha _{i\sigma }^ +  \alpha _{j\sigma }  - }
\nonumber \\
 -\sum\limits_{i \ne j} {\delta t_{ij}  < \alpha _{i\sigma }^ +  \alpha _{j\sigma }  > } 
\label{energy21} 
\end{eqnarray}
The first term of  ~(\ref{energy21}) is a nonoperator part and last two are connected with correlation corrections to superconducting state. In our mean field theory this corrections are not considered.

Therefore,  we have
\begin{eqnarray}
\Delta H = \Delta H_1, 
\label{energy22} 
\end{eqnarray}
where 
\[
\frac{1}{N}\sum\limits_k {\Delta _k^{2p} }  = \frac{{\Delta ^{2p} }}{{\pi ^2 }}\int\limits_0^\pi  {dx\int\limits_0^\pi  {dy(\cos x \pm \cos y)^{2p} } } 
\]
and
\begin{eqnarray}
\delta t_{ii}  =  - \frac{{\Delta ^2 }}{{2\tilde \mu }} + \frac{{9\Delta ^4 }}{{32\tilde \mu ^3 }} - \frac{{75\Delta ^6 }}{{192\tilde \mu ^5 }} + ...
\label{energy23} 
\end{eqnarray}

Finally, an unperturbed  BCS Hamiltonian in coordinate spase takes the simple  form 
\begin{eqnarray}
\hat H_{0f}  =  - \sum\limits_i {\mu _g {\kern 1pt} \alpha _{i\sigma }^ +  \alpha _{j\sigma } }, 
\label{energy24} 
\end{eqnarray}
where the renormalized chemical potential   $\mu_g$  looks as
\begin{eqnarray}
\mu _g  = \tilde \mu  - \delta t_{ii} 
\label{energy25} 
\end{eqnarray}

In view of the unitary transformed Hamiltonian ~(\ref{energy24}) one can build the perturbation theory with operator $\hat V$  from (6) in which $\tilde c_{i\sigma }^ +$ and $\tilde c_{j\sigma}$ replaced by ${\tilde {\tilde c}_{i \sigma }^ +}$  and ${\tilde {\tilde c}_{j \sigma }}$, respectively, where 

\begin{eqnarray}
\begin{array}{c}
\tilde {\tilde c}_{i\sigma }  = \frac{1}{N}\sum\limits_{{\bm k}p} {e^{i{\bm k}({\bm R}_i  - {\bm R}_p )} } \left\{ { u_{\bm k}^* \alpha _{p\sigma }  + v_{\bm k} \alpha _{p - \sigma }^ +  } \right\} \\ 
\tilde {\tilde c}_{i\sigma }^ +   = \frac{1}{N}\sum\limits_{{\bm k}p} {e^{ - i{\bm k}({\bm R}_i  -{\bm R}_p )} } \left\{ {u_{\bm k} \alpha _{p\sigma }^ +   + v_{\bm k}^* \alpha _{p - \sigma } } \right\} \\ 
\end{array}
\label{energy26}
\end{eqnarray}
Here  the coefficients of the u-v transform are
\begin{eqnarray}
u_{\bm k}  = \frac{{\tilde \mu  - \tilde E_{\bm k} }}{{\sqrt {2\tilde E_{\bm k} (\tilde E_{\bm k}  - \tilde \mu )} }},\quad v_{\bm k}  = \frac{{\Delta _{\bm k} }}{{\sqrt {2\tilde E_{\bm k} (\tilde E_{\bm k}  - \tilde \mu )} }}
\label{energy27}
\end{eqnarray}
\section{PERTURBATION THEORY FOR ELECTRON SYSTEM  IN  A   SUPERCONDUCTING  STATE}
The scattering matrix formalism for a system with strongly correlated electrons differs  from that in the band theory of metals. Indeed, in our case we must exclude the upper Hubbard's band by  Gutzwiller's projection operator. As a result one can not use the  wave representation at disentanglement of correlators  arising from series of  the perturbation theory. Unfortunately, the BCS Hamiltonian  is diagonalized  if  and only if  we work in $\bm k$-space. That's why, the coordinate representation ~(\ref{energy26}) is introduced  to connect the coordinate and $\bm k$-spaces of transformed Hamiltonian. There is a powerful method of Hubbard operators to account for excluding of double electron site occupancy. In works \cite{Izyum01,Izyum02,Zubov01} the general diagram method for Hubbard operator was presented. Let us dwell on the main statements related to our model.

Let us introduce the Hubbard's operators $X^{ik}  = \left| {\psi _i } \right\rangle \left\langle {\psi _k } \right|$, where we have three electron wave functions $\left| {\psi _0 } \right\rangle  = \left| 0 \right\rangle ,\,\left| {\psi _\sigma  } \right\rangle  = \left| \sigma  \right\rangle $, corresponding to hole, spin up $\\$ ( $\sigma$=+) and spin down ($\sigma$=-) electron states, respectively. Apaprently, in a normal state $\alpha _{p\sigma }  = X_p^{0\sigma } ,\;\alpha _{p\sigma }^ +   = X_p^{\sigma 0}$ and anticommutator 
$\left\{ {\alpha _{p\sigma } \alpha _{p\sigma }^ +  } \right\} = F_p^{\sigma 0}  = X_p^{\sigma {\kern 1pt} \sigma }  + X_p^{0{\kern 1pt} 0} $
 , which differs from unit as for destruction and creation operators. It is a result of neglecting the upper Hubbard's band. The main task is to find the average of operators $F_p^{\sigma 0}$  and $c_{i - \sigma } c_{j\sigma }$. In the first  case  we obtain the self-cosistent equation for chemical potential of the paramagnetic state. Apparently, from condition $\Delta /\mu  <  < 1$ one can neglect the influence of order parameter $\Delta$  on $\mu$ . In the second case we will have the equation for a gap function. Since the femion and boson subsystems are divided in accordance with Hamiltonian ~(\ref{energy05}) now we will consider creation and destruction operators with one tilda corresponding to u-v transform. The second tilda will reflect an unitary transform ~(\ref{energy07}). The average of operator A is determined by a standard manner:
\begin{eqnarray}
<\textit A>  = \frac{1}{{ < \sigma (\beta ) > _0 }} < A\sigma (\beta ) > _0, 
\label{energy28}
\end{eqnarray}
where the symbol   $\left\langle\right\rangle_0$=$Sp(\exp ( - \beta \hat H_0 ...)/Sp(\exp ( - \beta \hat H_0 ))$
 denotes a  statistical averaging over the unperturbed Hamiltonian with temperature 1/$\beta$ =$\textit{T}$, $\textit{Sp}$ is a trace of the operator. The $\textit{S}$ matrix  is written as \cite{Mahan}
\begin{eqnarray}
\sigma (\beta ) = \sum\limits_{n = 0}^\infty  {\frac{{( - 1)^n }}{{n!}}} \int\limits_0^\beta  {...\int\limits_0^\beta  {d\tau _1 ...d\tau _n T_\tau  \left\{ { V(\tau _1 )... V(\tau _n )} \right\}} }
\nonumber \\
\label{energy29}
\end{eqnarray} 

In expression for $\sigma(\beta)$  the symbols  $V(\tau _i ) = e^{\hat H_0 \tau _i } \hat Ve^{ - \hat H_0 \tau _i } $ and $\textit{T}_\tau$    are operator  in interaction representation and time-ordering operator, respectively. Now the task is to calculate all possible averages of the product of  \textit{A} operators for different sites. Using the Vick's theorem for Hubbard's operators this correlators can be reduced to product of  semi-invariants of  the diagonal operators and unperturbed Fermi Green's functions                                             $\textit{G}_{\alpha\beta}(\tau)=-< \textit{T}_\tau \textit{X}^{\alpha\beta}(\tau)  \textit{X}^{\beta\alpha}(0)>_0/<\textit{F}^{\sigma0}>_0$ \cite{Izyum01,Izyum02}.  The Fourier transform of these  functions has the form
\begin{eqnarray}
G_{\alpha \beta } (i\omega _n ) = \frac{1}{{2\beta }}\int\limits_{ - \beta }^\beta  {e^{ - i\omega _n \tau } G_{\alpha \beta } (\tau )d\tau }  = \frac{1}{\beta }\frac{1}{{i\omega _n  + \varepsilon _{\alpha\beta}  }},
\nonumber \\
\label{energy30}
\end{eqnarray} 
where $\omega _n  = (2n + 1)\pi /\beta $, $\varepsilon _{\alpha\beta}=\varepsilon _{\alpha}   - \varepsilon _{\beta  }$ and  $\epsilon_\alpha$  is the energy level of unperturbed Hamiltonian $\hat H_{0f}$. In the case where is account of the electron-phonon interaction the unperturbed Fermi-Bose Green's function is written as $\tilde G_{\alpha \,\beta } (\tau ) =  -  < T_\tau  X^{\alpha \,\beta } (\tau )Y(\tau )X^{\beta \,\alpha } (0)Y^ +  (0) > _0 /$   $< F^{\sigma 0}  > _0 $. The Fourier component $\tilde G_{\alpha \beta } (i\omega _n )$
  of  $\tilde G_{\alpha \,\beta } (\tau )$ is expressed as follows \cite{Zubov01}:
\begin{eqnarray}
\tilde G_{\alpha \,\beta } (i\omega _n )=\ \ \ \ \ \ \ \ \ \ \ \ \ \ \ \nonumber\\
= \frac{1}{\beta }f(\varepsilon _{\alpha \beta } )\sum\limits_{m =  - \infty }^{ + \infty } {d_m \frac{{e^{\beta \varepsilon _{\alpha \,\beta }  + \frac{1}{2}\beta m\omega _0 }  + e^{ - \frac{1}{2}\beta m\omega _0 } }}{{i\omega _n  + m\omega _0  + \varepsilon _{\alpha \beta } }}},
\nonumber \\
\label{energy31}
\end{eqnarray}
where $f(x) = 1/(e^{\beta x}  + 1\;)$ and $B = 1/(e^{\beta \omega _0 }  - 1)$ are the Fermi and Bose distributions, respectively, $d_m  = e^{ - \lambda ^2 (2B + 1)} I_m \left( {2\lambda ^2 \sqrt {B(B + 1)} } \right)$, $I_m (x)$ are the Bessel functions of complex argument and  $\omega_n$  is the same as in ~(\ref{energy30}). 

Unfortunately, the Vick's theorem cannot be used  for transformed Hubbard's operators in acoordance with ~(\ref{energy07}). That's why, for averaging we have to separate the boson subsystem from fermion. In case of isolated pairings a such separation is not needed. As it will be seen later a similar situation  is realized for effective kinematic interaction and diagrams to be formed by one effective line of interactions and one unperturbed Green's function. Further, we denote by  $U_{ep} (\tau _i  - \tau _j ) =  < T_\tau  Y_i (\tau _i )Y_i^ +  (\tau _j ) > _0 $

Let us write the possible pairings between creation and destruction operators. The normal operator pairing has the form
\begin{eqnarray} \label{energy32} T_{\tau } \underrightarrow{\tilde{\tilde{c}}_{i\sigma } (\tau )\tilde{\tilde{c}}_{j\sigma }^{+} (0)}=\frac{1}{N^{2} } \sum _{{\bm k}_{1} ,{\bm k}_{2} ,p}e^{i{\bm k}_{1} {\bm R}_{i} -i{\bm k}_{2} {\bm R}_{j} } e^{i({\bm k}_{1} -{\bm k}_{2} ){\bm R}_{p} }\cdot
\nonumber \\ 
\cdot\left\{-u_{{\bm k}_{1} }^{*} u_{{\bm k}_{2} } \tilde{G}_{0\sigma } (\tau )F_{p}^{\sigma 0} -v_{{\bm k}_{2} }^{*} v_{{\bm k}_{1} } \tilde{G}_{-\sigma 0} (\tau )F_{p}^{-\sigma 0} \right\}
\nonumber\\
\end{eqnarray} 
and  an average over the Hamiltonian $\hat{H}_{0f} $ gives the next unperturbed Green's functions:
\begin{eqnarray} 
\label{energy33}
\tilde{\tilde{G}}_{0\sigma } (\tau )=-<T_{\tau } \tilde{\tilde{c}}_{i\sigma } (\tau)\tilde{\tilde{c}}_{j\sigma }^+(0)>_{0} =\frac{1}{N} \sum\limits_{\bm k}e^{i\bm k(\bm R_{i} - \bm R_{j})}\cdot
\nonumber \\
\left(\left|u_{\bm k}\right|^{2} \tilde{G}_{0\sigma } (\tau )<F_{p}^{\sigma 0} >_{0} +\left|v_{\bm k} \right|^{2} \tilde{G}_{-\sigma 0} (\tau )<F_{p}^{-\sigma 0} >_{0} \right),
\nonumber \\\end{eqnarray}
where $\tilde{G}_{0\sigma } (i\omega _{n} )$ is unperturbed Green's function  from ~\eqref{energy31}. The anormal operator pairing is presented as
\begin{eqnarray} 
T_{\tau } \underrightarrow{\tilde{c}_{m\sigma } (\tau )\tilde{c}_{l-\sigma } (0)}=\frac{1}{N^{2} } \sum _{{\bm k}_{1} ,{\bm k}_{2} ,p}e^{i{\bm k}_{1} {\bm R}_{m} +i{\bm k}_{2} {\bm R}_{l} } e^{-i({\bm k}_{1} +{\bm k}_{2} ){\bm R}_{p} }\cdot
\nonumber \\
\cdot\left(-u_{{\bm k}_{1} }^{*} v_{{\bm k}_{2} } G_{0\sigma } (\tau )F_{p}^{\sigma 0} -v_{{\bm k}_{1} } u_{{\bm k}_{2} }^{*} G_{-\sigma 0} (\tau )F_{p}^{-\sigma 0} \right)\ \ \ \ \ 
\label{energy34}
\end{eqnarray}
Here the arrows are directed from ''active'' operator to ''passive'' with the use of  Vick's theorem. In particular, the average of  ~(\ref{energy34}) at $\tau\rightarrow-0$ gives unperturbed gap functions:
\begin{eqnarray}  
<\tilde{c}_{l-\sigma } \tilde{c}_{m\sigma } >_{0} =-\frac{1}{N} \sum _{{\bm k}}e^{i{\bm k}({\bm R}_{l} -{\bm R}_{m} )} u_{-{\bm k}}^{*} v_{{\bm k}} <F^{\sigma 0} >_{0}  =
\nonumber \\
=\frac{1}{N} \sum _{{\bm k}}e^{i{\bm k}({\bm R}_{l} -{\bm R}_{m} )} \varphi _{{\bm k}\sigma }^{11(0)}(0)\ \ \ \ \ 
\label{energy35}  
\end{eqnarray} 

The pairing with diagonal operator takes the form :
\[T_{\tau } \underrightarrow{\tilde{c}_{i\sigma } (\tau )F_{m}^{\sigma 0} }=-\frac{1}{N} \sum\limits_{\bm k}e^{i\bm k(\bm R_{i}-\bm R_{m} )} v_{{\bm k}} \tilde{G}_{-\sigma 0} (\tau ) \alpha _{m-\sigma }^{+} \] 
The similar expressions can be presented for other pairings. One can point out that anticommutator
\[\begin{array}{c} 
\left\{\tilde c_{i\sigma } \tilde c_{i\sigma }^ + \right\}  = \tilde F_i^{\sigma 0}  = \frac{1}{N^2}\sum\limits_{\bm k_1 \bm k_2 p}e^{i(\bm k_1  - \bm k_2 )(\bm R_i  - \bm R_p )}\cdot \\
\cdot\left( u_{\bm k_1 }^* u_{\bm k_2 } F_p^{\sigma 0}  + v_{\bm k_1 } v_{\bm k_2 }^* F_p^{ - \sigma 0} \right)\end{array}
\]

We shall calculate the average  value $<\tilde{c}_{l-\sigma } \tilde{c}_{m\sigma } >$.This average is conveniently calculated by replacement  
\[<\tilde{c}_{l-\sigma } \tilde{c}_{m\sigma } >=\mathop{\lim }\limits_{\tau \to +0} <T_{\tau } \tilde{c}_{l-\sigma } (\tau )\tilde{c}_{m\sigma } (0)>\] 

Let $B_{0\sigma } (\tau _j  - \tau _i ,\bm k)$
 is the Fourier components of the effective kinematic interaction caused by perturbation $\hat V$ from ~(\ref{energy03}). Then it is necessary to find an average of the next pairings in the first approximation of the perturbation theory:
\begin{eqnarray}
\begin{array}{c} 
-\int _{0}^{\beta }d\tau _{j} \int _{0}^{\beta }d\tau _{i} \frac{1}{N} \sum\limits_{ij{\bm k}}\frac{1}{\beta } B_{0\sigma } (\tau _{j} -\tau _{i} ,{\bm k})e^{i{\bm k}({\bm R}_{j} -{\bm R}_{i} )}\cdot \\
\cdot<T_{\tau } \overrightarrow{\tilde{c}_{l-\sigma } (\tau )\tilde{c}_{m\sigma } (0)}\underleftarrow{\tilde{\tilde{c}}_{j\sigma }^{+} (\tau _{j} )\tilde{\tilde{c}}_{i\sigma } (\tau _{i} )}>_{0}    = \\ {\; \; \;} \quad = - \frac{1}{N}\sum\limits_{\bm k}e^{i{\bm k}({\bm R}_l  - {\bm R}_m )}\cdot \\
\cdot ( u_{\bm k}^* v_{ - \bm k} G_{0 - \sigma } (\tau )\beta \delta \tilde {\tilde\mu}_{- \sigma }  + u_{ - \bm k}^* v_{\bm k} G_{\sigma 0} (\tau )\beta \delta \tilde {\tilde\mu}_\sigma )\\\\\\\\ 
\end{array}
\label{energy36}
\end{eqnarray}
where function   $\beta \delta {\kern 1pt} \tilde{\tilde{\mu }}_{\sigma } $ is determined as
\begin{equation}
\label{energy37}
\beta \delta \tilde{\tilde{\mu }}_{\sigma } =\frac{1}{N} \sum _{{\it q}\omega _{{\it n}} }\beta B_{0\sigma } ({\it q},i\omega _{n} )\tilde{\tilde{G}}_{0\sigma } (i\omega _{n} )  \end{equation} 
The high orders of perturbation theory with pairings of type ~\eqref{energy36} renormalize the combined occupancy, $<F^{\sigma 0} >_{0} $ to $<F^{\sigma 0} >_{1} $ in accordance with expanding  function  
\[<F^{\sigma 0} (\varepsilon _{\sigma } ,\varepsilon _{-\sigma } ,\lambda )>_{0} =\frac{e^{-\beta \varepsilon _{\sigma } } +e^{-\beta (\lambda +\varepsilon _{\sigma } +\varepsilon _{-\sigma } )} }{e^{-\beta \varepsilon _{\sigma } } +e^{-\beta (\lambda +\varepsilon _{\sigma } +\varepsilon _{-\sigma } )} +1} \]
to be equal to $<F^{\sigma 0} >_{0} $  at $\lambda {\rm =}-\varepsilon _{-} -\varepsilon _{+} $ into a Taylor's series \cite{Zubov01}
\begin{eqnarray} 
\label{energy38} 
\begin{array}{c} 
<F^{\sigma 0} >_{1} =<F^{\sigma 0} >_{0} -\beta \delta \tilde{\tilde{\mu }}_{\sigma } \partial _{\sigma } <F^{\sigma 0} >_{0}-\\
-\beta \delta \tilde{\tilde{\mu }}_{-\sigma }\partial _{-\sigma } <F^{\sigma 0} >_{0}+ \frac{1}{2!} (\beta \delta \tilde{\tilde{\mu }}_{\sigma })^{2} \partial _{\sigma }^{2}<F^{\sigma 0} >_{0}+
\\ +\frac{1}{2!} (\beta \delta \tilde{\tilde{\mu }}_{-\sigma } )^{2} \partial _{-\sigma }^{2} <F^{\sigma 0} >_{0} -...= \\=<F^{\sigma 0} (\varepsilon _{\sigma } +\delta \tilde{\tilde{\mu }}_{\sigma } ,\varepsilon _{-\sigma } +\delta \tilde{\tilde{\mu }}_{-\sigma }^{} ,-\varepsilon _{\sigma } -\varepsilon _{-\sigma } )>_{0} 
\end{array} 
\end{eqnarray}
Thus, on summing ~\eqref{energy35},  ~\eqref{energy36} and so on, we  obtain
\begin{eqnarray} 
\label{energy39}
\varphi _{{\bm k}\sigma }^{11} (\tau \to +0)=u_{{\bm k}}^{*} v_{-{\bm k}} G_{0-\sigma } (\tau )<F^{-\sigma 0} >_{1}+\nonumber \\ 
+u_{-{\bm k}}^{*} v_{{\bm k}} G_{\sigma 0} (\tau )<F^{\sigma 0} >_{1}  
\end{eqnarray}

The next more comlicated pairings appear as
\begin{widetext}
\[\begin{array}{c}
{-\int _{0}^{\beta }d\tau _{j} \int _{0}^{\beta }d\tau _{i} \frac{1}{N} \sum\limits_{ij{\bm k}}B_{0\sigma } (\tau _{j} -\tau _{i} ,{\bm k})e^{i{\it k}({\bm R}_{j} -{\bm R}_{i} )} <T_{\tau } \overleftarrow{\tilde{c}_{l-\sigma } (\tau )\underleftarrow{\tilde{c}_{m\sigma } (0)\tilde{\tilde{c}}_{j\sigma }^{+} (\tau _{j} )\tilde{\tilde{c}}_{i\sigma } (\tau _{i} )}}>_{0}    =} \\ {-\int _{0}^{\beta }d\tau _{j} \int _{0}^{\beta }d\tau _{i} \frac{1}{N} \sum\limits _{ij{\bm k}}B_{0\sigma } (\tau _{j} -\tau _{i} ,{\bm k})e^{i{\bm k}({\bm R}_{j} -{\bm R}_{i} )} <T_{\tau } \underleftarrow{\tilde{c}_{l-\sigma } (\tau )\tilde{c}_{i\sigma } (\tau _{i} )}\tilde{c}_{m\sigma } (0)\tilde{c}_{j\sigma }^{+} (\tau _{j} )>_{0} \cdot    U_{ep} (\tau _{j} -\tau _{i} )=} \\ {=\frac{1}{N} \sum\limits _{\bm k}e^{i{\bm k}({\bm R}_{l} -{\bm R}_{m} )} \varphi _{{\bm k}\sigma }^{12} (\tau ) },
\end{array}  \]
\end{widetext}
where 
\begin{widetext}
\[\begin{array}{c} 
{\varphi _{{\bm k}\sigma }^{12} (\tau \to +0)=\int _{0}^{\beta }d\tau _{j} \int _{0}^{\beta }d\tau _{i} \frac{1}{N} \sum\limits_{{\bm k}'}\frac{1}{\beta } B_{0\sigma } (\tau _{j} -\tau _{i} ,{\bm k}')   \cdot } \\ {\quad \quad \cdot \left\{\begin{array}{l} {u_{{\bm k}'}^{{\it *}} v_{{\bm k}'}^{{\it *}} v_{{\bm k}}^{{\it 2}} G_{0\sigma }^{} (\tau _{i} -\tau )G_{0-\sigma } (\tau _{j} -\tau )G_{-\sigma 0}^{} (-\tau )<F^{-\sigma 0} >_{0} -} \\ {v_{{\bm k}'} u_{{\bm k}'} (u_{{\bm k}}^{{\it *}} )^{2} G_{-\sigma 0}^{} (\tau _{i} -\tau )G_{\sigma 0} (\tau _{j} -\tau )G_{0\sigma }^{} (-\tau )<F^{\sigma 0} >_{0} } \end{array}\right\}\cdot U_{ep} (\tau _{i} -\tau _{j} )-} \\ {-\int _{0}^{\beta }d\tau _{j} \int _{0}^{\beta }d\tau _{i} \frac{1}{N} \sum\limits_{{\bm k}'}\frac{1}{\beta } B_{0\sigma } (\tau _{j} -\tau _{i} ,{\bm k}')   U_{ep} (\tau _{i} -\tau _{j} )\cdot \left\{\begin{array}{l} {\left|u_{{\bm k}'} \right|^{2} v_{{\bm k}} u_{{-\bm k}}^{{\it *}} G_{0\sigma }^{} (\tau _{i} -\tau )G_{\sigma 0} (\tau _{j} )\partial _{\sigma } <F^{\sigma 0} >_{0} +} \\ {+u_{{\bm k}'}^{{\it *}} v_{{\bm k}'}^{{\it *}} v_{{\bm k}} v_{{-\bm k}} G_{0\sigma }^{} (\tau _{i} -\tau )G_{0-\sigma } (\tau _{j} )\partial _{-\sigma } <F^{\sigma 0} >_{0} +} \\ {+v_{{\bm k}'} u_{{\bm k}'} u_{{\bm k}}^{{\it *}} u_{{-\bm k}}^{{\it *}} G_{-\sigma 0}^{} (\tau _{i} -\tau )G_{\sigma 0} (\tau _{j} )\partial _{-\sigma } <F^{\sigma 0} >_{0} +} \\ {+\left|v_{{\bm k}'} \right|^{2} u_{{\bm k}}^{{\it *}} v_{{-\bm k}} G_{-\sigma 0}^{} (\tau _{i} -\tau )G_{0-\sigma } (\tau _{j} )\partial _{-\sigma } <F^{\sigma 0} >_{0} } \end{array}\right\}} 
\end{array}\]
\end{widetext} 
Here, the cumulants $<F_{p}^{\sigma 0} F_{p'}^{\sigma 0} >_{0} $ and $<F_{p}^{-\sigma 0} F_{p'}^{\sigma 0} >_{0} $ correspond to linked diagrams and expressed in the terms of  derivatives $\partial _{\sigma } <F^{\sigma 0} >_{0} $ and $\partial _{-\sigma } <F^{\sigma 0} >_{0} $, respectively, where $\partial _{\sigma } =\partial /\partial (-\beta {\kern 1pt} \varepsilon _{\sigma } ).$

Also, we have to evaluate the mean value $K_2=\left\langle T_{\tau }\tilde{c}_{l-\sigma}(\tau)\tilde{c}_{m\sigma}(0)\tilde{\tilde{c}}_{j-\sigma }^{+} (\tau_{j})\tilde{\tilde{c}}_{i-\sigma }(\tau _{i})\right\rangle_{0}$ at $\tau \to +0$. If we make in previous correlator $K_1=\left\langle T_{\tau }\tilde{c}_{l-\sigma }(\tau )\tilde{c}_{m\sigma} (0)\tilde{\tilde{c}}_{j\sigma }^{+}(\tau _{j})\tilde{\tilde{c}}_{i\sigma } (\tau _{i})\right\rangle_{0}$ the replacement $\sigma \to -\sigma$ and  $m\leftrightarrow l$, we obtain $-\textit{K}_2$ in the limit for $\tau  \to  - 0$
Thus,  it follows from $\textit{K}_1$ :
\[\varphi _{{\bm k}\sigma }^{21} (0)=-\varphi _{-{\bm k}-\sigma }^{12} (\tau \to -0)\]
One can point out that  Fourier transform of $<\tilde{c}_{l-\sigma } \tilde{c}_{m\sigma } >$ is $\varphi _{{\it k}\sigma } $, i.e.
\[<\tilde{c}_{l-\sigma } \tilde{c}_{m\sigma } >=\frac{1}{N} \sum _{\bm k}e^{i{\bm k}({\bm R}_{l} -{\bm R}_{m} )} \varphi _{{\bm k}\sigma }  \]
It follows that $<\tilde{c}_{-k-\sigma } \tilde{c}_{k\sigma } >$=$\varphi _{-{\it k}\sigma } $. Using the relations $G_{0\sigma } (\tau \to +0)=G_{\sigma 0} (\tau \to -0)=0$ and $G_{\sigma 0} (\tau \to +0)=G_{0\sigma } (\tau \to -0)=-1$ at $\varepsilon _\sigma   < 0$
 as well as that in paramagnetic phase the  indices $\sigma$ and  -$\sigma$ denote the same state  we obtain 
\begin{widetext}
\begin{eqnarray} 
\label{energy40} 
\begin{array}{c} {\varphi _{{\bm k}\sigma }^{12} (0)+\varphi _{{\bm k}\sigma }^{21} (0)=\int _{0}^{\beta }d\tau _{j} \int _{0}^{\beta }d\tau _{i} \frac{1}{N} \sum\limits_{{\bm k}'}\frac{1}{\beta } B_{0\sigma } (\tau _{j} -\tau _{i} ,{\bm k}') U_{ep} (\tau _{i} -\tau _{j} )<F^{\sigma 0} >_{0} \cdot } \\ {\; \; \quad \quad \cdot \left\{v_{{\bm k}'} u_{{\bm k}'} (u_{{\bm k}}^{{\it *}} )^{2} G_{-\sigma 0}^{} (\tau _{i} )G_{\sigma 0} (\tau _{j} )+u_{{\bm k}'}^{{\it *}} v_{{\bm k}'}^{{\it *}} v_{{\bm -k}}^{{\it 2}} G_{0-\sigma }^{} (\tau _{i} )G_{0\sigma } (\tau _{j} )\right\}} 
\end{array} 
\end{eqnarray}
\end{widetext}
Apparently, that integrals
\begin{widetext}
\[\begin{array}{c} \int _{0}^{\beta }d\tau _{j} \int _{0}^{\beta }d\tau _{i} \frac{1}{N} \sum\limits_{\bm k'}\frac{1}{\beta } B_{0\sigma } (\tau _{j} -\tau _{i} ,\bm k')U_{ep} (\tau _{i} -\tau _{j} )G_{-\sigma 0}(\tau _{i} )G_{\sigma 0} (\tau _{j} )=\\ =\int _{0}^{\beta }d\tau _{j} \int _{0}^{\beta }d\tau _{i} \frac{1}{N} \sum\limits_{\bm k'}\frac{1}{\beta } B_{0\sigma } (\tau _{j} -\tau _{i} ,{\bm k}')U_{ep} (\tau _{i} -\tau _{j} )G_{0-\sigma }^{} (\tau _{i} )G_{0\sigma } (\tau _{j} )= \\ =\frac{1}{N} \sum\limits_{\bm k' n_{1} n_{2} }\frac{1}{\beta } B_{0\sigma } (i\omega _{n_{1}} ,\bm k')U_{ep} (i\omega _{n_{1} } -i\omega _{n_{2} } ) G_{-\sigma 0}(i\omega _{n_{2} } )G_{\sigma 0} (-i\omega _{n_{2} } ) \end{array}\]
\end{widetext}
Here we used  the next property of the unperturbed Green's function:
\[G_{0\sigma }^{} (i\omega _{n} )=-G_{\sigma 0} (-i\omega _{n} )\]
Thus, the left hand of Eq.~\eqref{energy40} does not depend on wave vector \textbf{k}  because of  relations ~\eqref{energy27} and this contribution in gap function ~\eqref{energy11} is equal zero. Then we have for gap function
\[<\tilde{c}_{-\bm k-\sigma } \tilde{c}_{\bm k\sigma } >=-u_{{\bm -k}}^{*} v_{{\bm k}} <F^{\sigma 0} >_{1} \] 
Using Eqs.~\eqref{energy27} we obtain the equation for gap
\begin{eqnarray}
\label{energy41} 
\Delta _{{\bm k}} =\frac{1}{N} \sum _{{\bm q}}J({\bm q}+{\bm k})\frac{\Delta _{{\bm q}} }{{\rm 2}\left|\tilde{{\it E}}_{{\bm q}} \right|}  <F^{\sigma 0} >_{1}\ \ \  
\end{eqnarray}
This equation in the limit T-$>$0 coincides practically with a similar equation obtained by G.Baskaran Z. Zou and  P.W. Anderson in Ref. \cite{Baskar}. There is one essential difference. In Ref. \cite{Baskar} the temperature factor depends on wave vector and tends to 1 at \textit{T}-$>$0. In our case the electron-hole presence is accounted for. As a result we have \textbf{k} --independent factor $<F^{\sigma 0} >_{1} $  to be equal to 1/2 at temperature T-$>$0. This factor is determined by Eq.\eqref{energy38}. Here we can put $\delta \tilde{\tilde{\mu }}_{\sigma } =\delta \tilde{\mu }_{\sigma } $when finding the critical temperature of superconducting state.
Then one can write
\begin{eqnarray}
\label{energy42}
< F^{\sigma 0}  > _1  = \frac{{e^{\beta E_\sigma  }  + 1}}{{1 + e^{\beta E_\sigma  }  + e^{\beta E_{ - \sigma } } }},
\end{eqnarray}
where  $E_{\sigma } =-\varepsilon _{\sigma } +\delta \tilde{\mu }_{-\sigma }^{} $ and
\begin{eqnarray} 
\label{energy43} 
\beta \delta \tilde{\mu }_{\sigma } =\frac{1}{N} \sum _{{\bm q}\omega _{{\it n}} }\beta B_{0\sigma } ({\bm q},i\omega _{n} )\tilde{G}_{0\sigma } (i\omega _{n} )  
\end{eqnarray} 
with $\tilde{G}_{0\sigma } (i\omega _{n} )$ from ~\eqref{energy31}. The solution of ~\eqref{energy41} at  temperature \textit{T}$_{\textit{SC}}$ when $\Delta$=0 gives the self-consistent equation for temperature of the superconducting transition:
\begin{eqnarray} 
\label{energy44} 
T_{SC} =\frac{\tilde{\mu }+\delta \, \tilde{\mu }_{\sigma } }{\ln \frac{J-2\tilde{\mu }}{4\tilde{\mu }-J} }  
\end{eqnarray}
From \eqref{energy44} follows the next requirement on the chemical potential:
\begin{eqnarray} 
\label{energy45} J/4\le \tilde{\mu }\le J/3 
\end{eqnarray}
This condition is rigorous and that's why the preceding  spin-fluctuation theories are failed in the explanation  of  high-temperature superconductivity. As will be seen from a next section the chemical potential of  paramagnetic state of the strongly correlated electrons  substantially exceed an exchange parameter \textit{J}, i.e. the strong charge-spin fluctuations destroy the Cooper's pairs.

\section{NORMAL STATE OF ELECTRONS  IN THE ABSENCE OF  ELECTRON-PHONON INTERACTION}

\begin{figure}[h]
\begin{center}
\includegraphics[width=\columnwidth]{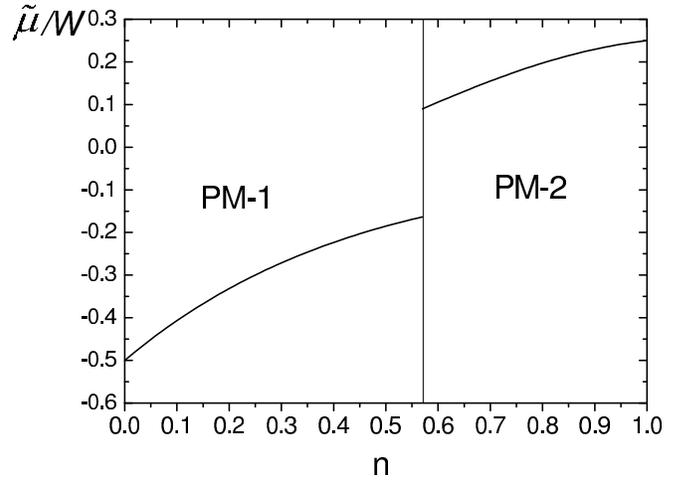}
\end{center}
\caption{The concentration dependence of the chemical potential in PM-1 and PM-2 phases in units of  bandwidth  \textit{W}. At  \textit{n}=0.5714 there is a disrupt of $\tilde{\mu}$}.
\label{fig.1}
\end{figure}

The theory of effective self-consistent field and phase transition in a system of the strongly correlated  d- electrons of cuprates was developed  by us in  works \cite{Zubov02,Myron}. In particular, the equation for chemical potntial in paramagnetic state is written as
\begin{eqnarray} 
\label{energy46} 
<F^{\sigma 0} >=1-\frac{n}{2} =<F^{\sigma 0} >_{1} -\tilde{\nu }_{^{_{-\sigma } } } <F^{-\sigma 0} >_{1},                                        
\end{eqnarray}
where $<F^{\sigma 0} >_{1} $ is determined by Eq.\eqref{energy42}  and  at \textit{g}=0
\begin{eqnarray} 
\label{energy47} 
\delta \tilde{\mu }_{\sigma }^{} =\delta \mu _{\sigma }^{} =\frac{1}{N} \sum _{{\it q}}{\it t(}{\bm q)} f(E_{{\bm q}\sigma }^{} ) 
\end{eqnarray}
\begin{eqnarray} 
\label{energy48} 
\tilde{\nu }_{\sigma }^{} =\nu _{\sigma }^{} =\frac{1}{<F^{\sigma 0} >} \left\{\frac{1}{N} \sum _{{\bm q}}f(E_{{\bm q}\sigma }^{} )-f(\varepsilon _{\sigma } ) \right\} 
\end{eqnarray}
Here, the band energy $E_{{\bm q}\sigma }^{} =\varepsilon _{\sigma } +t({\bm q})<F^{\sigma 0} >$ and Fourier components of hoping integral \textbf{$t(\bm q)=\sum\limits_{ij}t_{ij} e^{-i{\bm q}({\bm r}_{i} -{\bm r}_{j} )} =2t(\cos (q_{x} a)+\cos (q_{y} a)) $ }for rectangular lattice with constant $\textit{a}$. Let the function $I(x)$  be given  by formula:
\begin{eqnarray}
\label{energy49}
I(x)=\int _{-2}^{x}D_{C} (x)dx,
\end{eqnarray}
where the electron density of state $D_{C} (x)$ has a form for rectangular lattice
\begin{eqnarray} 
\label{energy50} 
D_{C} (x)=\frac{1}{\pi ^{2} } K\left(\sqrt{1-\left(x/2\right)^{2} } \right) 
\end{eqnarray}
and  $\textit{K(x)}$ is a complete elliptic integral of the first order. At $\textit{T}=0$ and $\tilde{\mu }+\delta \mu _{\sigma }^{} >0$ it is easy to write the solution of  Eq.\eqref{energy46}  for chemical potential of paramagnetic (PM-2) phase:
\begin{eqnarray}
\label{energy51}
\tilde{\mu }/W=\frac{2-n}{8} I^{-1} \left(1-\frac{1}{2} (1-n)(2-n)\right),                      \end{eqnarray}
where $I^{-1}(x)$ is an inverse function of $\textit{I(x)}$. It coresponds to gas limit in a hole concentration $\textit{1-n}$, when  $<F^{\sigma 0} >_{1} =1/2$. Indeed, $<F^{\sigma 0} >=1-n/2$, i.e. at $n\sim 1$ we have $<F^{\sigma 0} >\sim 1/2$.  At $\textit{T}=0$ and $\tilde{\mu }+\delta \mu _{\sigma }^{} <0$ we obtain for paramagnetic (PM-1) phase:
\[\tilde{\mu }_{PM1} /W=\frac{2-n}{8} I^{-1} \left(\frac{n}{2} \left(1-\frac{n}{2} \right)\right),\] 
that corresponds to gas limit in electron concentration \textit{n}: $<F^{\sigma 0} >_{1} =1$ and $<F^{\sigma 0} >\sim 1$. In Fig.~\ref{fig.1}    the concentration dependence of the chemical potential in units of  bandwidth \textit{W} in PM-1 and PM-2 phases is presented. One can see the disrupt of $\tilde{\mu }$ at $n=n_{cr.}$= 0.5714. From Fig.1 it is easy to see  the correlation narrowing of band \textit{W} in PM-2 phase. Indeed, at $n=1$ we have $\tilde{\mu }/W$=0.25 that it less then $\tilde{\mu }/W$=0.5 for ferromagnetic state when a such narrowing is absent \cite{Zubov03}. In work \cite{Brinkman} a similar  narrowing of  PM is also observed. Unfortunately, this narrowing is unsuffient to fulfill the condition ~\eqref{energy45} even at $n=n_{cr.}$ when  $\tilde{\mu }(n_{cr.})/W\approx$ 0.09. With an increase in temperature the chemical potential is also increased.

\begin{figure}[h]
\begin{center}
\includegraphics[width=\columnwidth]{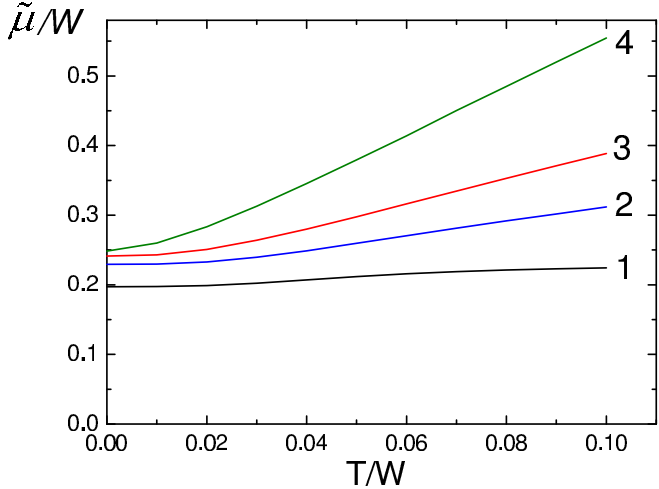}
\end{center}
\caption{The temperature dependencies of the chemical  potential at electron concentration $\textit{n}$=0.8, 0.9, 0.95 and 0.99 (curves 1-4, respectively).}
\label{fig.2}
\end{figure}

In Fig.~\ref{fig.2} the temperature dependencies of the chemical  potential at electron concentration $\textit{n}$=0.8, 0.9, 0.95 and 0.99 ( curves 1-4, respectively) are presented. These curves were obtained by numerical  solving of  the Eq.(~\ref{energy46}). It turns out that the inclusion of electron-phonon interaction may enforce essentially the correlation narrowing of band at which the conditions (\ref{energy45}) can be fulfilled.

\section{NORMAL STATE OF THE CUPRATE d-ELECTRONS WITH POLARON EXCITATIONS}

The problem  of  the polaron excitations in a  system of d-electrons was considered by many authors \cite{Kulic, Ciuchi, Alexandr, Edwards}. We will not analyze these works in detail but point out on the main their  limitations. Unfortunately, the authors  simplify the Hamiltonian $\hat V$    renormalizing  a hopping integral  $\textit{t}$ by factor \textbf{$e^{-\lambda ^{2} } $.} As a result we obtain a drastic decrease in temperature of the superconducting state \cite{Alexandr} and authors would have to use the effects of interactions of more  high order of smallness. In fact the situation is more complicated. It is connected with properties of the unperturbed Green's functions $\tilde{G}_{\alpha \, \beta } (i\omega _{n} )$. In series expansion of  Eq. (\ref{energy31}) for $\textit{m}$-th order we have product $e^{-\lambda ^{2} } $ and $\lambda ^{2m} /m!$. As was prooved by   G.D. Mahan in book \cite{Mahan} there is a Gaussian $\frac{1}{\sqrt{2\pi m} } \exp \left[-(\lambda ^{2} -m)/(2m)\right]$ instead of $e^{-\lambda ^{2} } $.Thus, with increasing $\lambda $ a number $\textit{m}$  of the polaron band, where the spectral function has a maximum,  is increased.

\begin{figure}[h]
\begin{center}
\includegraphics[width=\columnwidth]{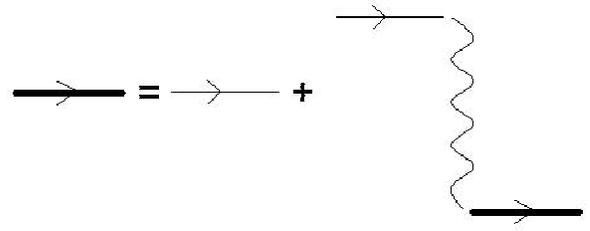}
\end{center}
\caption{Graphic equation for Green's function  $\Lambda _{0\sigma } (i\omega _{n} ,{\it q})$.}
\label{fig.3}
\end{figure}

Let us consider this question more thoroughly. In zeroth order of effective field for total Green's function   $\Lambda _{0\sigma } (i\omega _{n} ,{\bm q})$   the  graphic equation  is presented in Fig.~\ref{fig.3}, where the bold, thin stright   and wave lines  correspond  to $\beta {\kern 1pt} \Lambda _{0\sigma } (i\omega _{n} ,{\bm q})$, $\beta {\kern 1pt} \tilde{G}_{0\sigma } (i\omega _{n} )$ and $t(\bm q)$, respectively. The solution of this equation  is written as
\begin{eqnarray} 
\label{energy52} 
\beta {\kern 1pt} \Lambda _{0\sigma } (i\omega _{n} ,\bm q)=\frac{\beta {\kern 1pt} \tilde{G}_{0\sigma } (i\omega _{n} )<F^{\sigma 0} >}{1-\beta t(\bm q)\tilde{G}_{0\sigma } (i\omega _{n} )<F^{\sigma 0} >}  
\end{eqnarray}
In this equation we have  replaced $<F^{\sigma 0} >_{0} $ on the total average $<F^{\sigma 0} >$ to obtain the self-consistent parameter for effective kinematic field.  The main problem in Eq.(\ref{energy52}) is connected with determination of poles which are defined by equation
\begin{eqnarray} 
\label{energy53} 
1-\beta t(\bm q)\tilde{G}_{0\sigma } (i\omega _{n} )<F^{\sigma 0} >=0 
\end{eqnarray}

\begin{figure}[h]
\begin{center}
\includegraphics[width=\columnwidth]{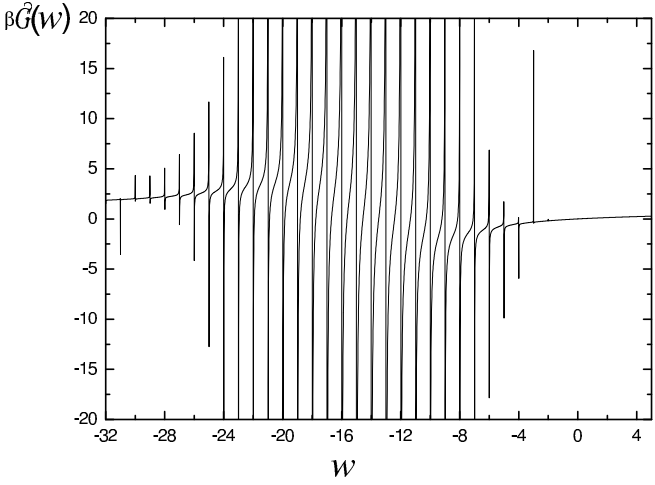}
\end{center}
\caption{The frequency dependence of the unperturbed fermion-boson Green's function $\beta \tilde{G}_{0\sigma } (w)$ at \textit{T}=0, \textit{g/W}=0.07, $\omega$0/\textit{W}=0.01875 and  $\tilde{\mu }$$>$0.}
\label{fig.4}
\end{figure}

Indeed, in diagram methods one must often evaluate  frequency summations. The traditional methods solve this problem if the poles of Matsubara Green's functions are known \cite{Mahan}. Unfortunately, Eq.(\ref{energy53})  for $i\omega _{n} $ gives the algebraic equation of  infinity order  and the task becomes unsolved. It turn out that one can overcome  this difficulty  by method of inverse function. To understand the  essence of a question we will simplify the Green's function $\tilde{G}_{0\sigma } (i\omega _{n} )$. It is supposed that in this system studied the condition $\omega_0/T>>1$  has to be. Then Bose factor  $\textit{B}$ in Eq.(\ref{energy31}) is replaced  by $exp(-\beta\omega_0)$ that allows to write $\tilde{G}_{0\sigma } (\Omega )$ in more simple form:
\begin{eqnarray}
\label{energy54} 
\tilde{G}_{0\sigma } (\Omega )=\kern 70pt         \nonumber\\
\frac{e^{-\lambda ^{2} } }{\beta } \sum _{m=0}^{\infty }\frac{\lambda ^{2m} }{m!} \left\{\frac{f(\varepsilon _{\sigma } )}{\Omega -\varepsilon _{\sigma } +m\omega _{0} }+\frac{1-f(\varepsilon _{\sigma } )}{\Omega -\varepsilon _{\sigma } -m\omega _{0} } \right\}
\end{eqnarray}
Let $(\Omega -\varepsilon _{\sigma } )/\omega _{0} =w$. Then we have
\begin{eqnarray}
\begin{array}{c}
\label{energy55} 
\beta \tilde{G}_{0\sigma } (w)=\frac{1}{w\omega _{0} } (M(1,1+w,-\lambda ^{2})f(\varepsilon _{\sigma } )+\\
+M(1,1-w,-\lambda ^{2} )(1-f(\varepsilon _{\sigma } )),
\end{array}
\end{eqnarray}
where \textit{ M(a,b,z)} is the confluent hypergeometric function of Kummer \cite{Abram}.  In Fig.~\ref{fig.4} the $\beta \tilde{G}_{0\sigma } (w)$ as function of $w$ is presented at \textit{T}=0 when $\varepsilon _{\sigma } =-\tilde{\mu }<0$, \textit{g/W}=0.07 and  phonon frequency $\omega_0$/\textit{W}=0.01875. As illustrated in Fig.~\ref{fig.4} this function  has monotonous behaviour between poles \textit{m} and \textit{m}+1. It allows to find the inverse function$[\beta \tilde{G}_{0\sigma } (\Omega )]^{-1} $ in  this area. Let us denote by $E_{n\bm q\sigma }$ the n-th root of the Eq.(\ref{energy53}). In the vicinity of  $E_{n\bm q\sigma }$  one can  expand $\beta \tilde{G}_{0\sigma } (\Omega )$ in powers $\Omega$-$E_{n\bm q\sigma }$  and  we have
\[\begin{array}{c}
 \beta {\kern 1pt} \Lambda _{0\sigma } (\Omega ,\bm q) \approx  \\ 
  \approx \sum\limits_n {\frac{{\beta \tilde G_{0\sigma } (E_{n\bm q\sigma } ) < F^{\sigma 0}  > }}{{ - t(\bm q) < F^{\sigma 0}  > \left. {\frac{{d\beta \tilde G_{0\sigma } (\Omega )}}{{d\Omega }}} \right|_{\Omega  = E_{n \bm q\sigma } } (\Omega  - E_{n\bm q\sigma } )}}}
 \end{array}\]

After  analytic continuation $\Omega$-$>$$\Omega$+i$\delta$ we obtain the imaginary part of the $\beta {\kern 1pt} \Lambda _{0\sigma } (\Omega ,{\bm q})$:
\[Im(\beta {\kern 1pt} \Lambda _{0\sigma } (\Omega ,{\bm q}))=\pi \sum _{n}\frac{\beta \tilde{G}_{0\sigma } (E_{n{\bm q}\sigma } )}{t({\bm q})\frac{d\beta \tilde{G}_{0\sigma } (\Omega )}{d\Omega } } \delta (\Omega -E_{n{\bm q}\sigma } ),\]  
where $\delta (x)$ is the Dirac delta function. The uniform spectral density \textit{$R_\sigma$}($\Omega$,0) is determined as
\begin{eqnarray} 
\label{energy56} 
R_{\sigma } (\Omega ,0)=-\frac{2}{N} \sum _{\bm q}Im(\beta {\kern 1pt} \Lambda _{0\sigma } (\Omega ,{\bm q}))  
\end{eqnarray}

To make the sum over $\bm q$ we wiil consider the dimensionless functions $\beta W\tilde{G}_{0\sigma } (\Omega )=F(\tilde{\Omega })$, $\tilde{E}_{n{\bm q}\sigma } =E_{n{\bm q}\sigma } /W$,  $\tilde{\Omega }=\Omega /W$ and $t({\bm q})/W=\varepsilon /4$, where the variable $\varepsilon \in [-2,2]$. Then the two-dimensional integral (\ref{energy56}) is replaced by one-dimensional  with density of state $D_{C} (\varepsilon )$ from (\ref{energy50}). The complicated delta function can be simplified by the relation \cite{Mahan}:
\[W\delta [\Omega -E_{n{\bm q}\sigma } ]=\frac{\delta (\varepsilon -\varepsilon _{0n} )}{\left|(\tilde{\Omega }-\tilde{E}_{n{\bm q}\sigma } )^{'} \right|_{\varepsilon =\varepsilon _{0n} } } ,\] 
where $\varepsilon _{0n} =\frac{4}{F(\tilde{\Omega })<F^{\sigma 0} >} $ is the function of the external frequency  $\tilde{\Omega }$. To find the derivative of $\tilde{E}_{n{\bm q}\sigma } $ as a function of variable $\epsilon$ we use the equation for \textit{n}-th pole
\[1-\frac{1}{4} \varepsilon <F^{\sigma 0} >F(\tilde{E}_{n\bm q\sigma } )=0\]
that gives    $\tilde{E}_{n\bm q\sigma } (\varepsilon )=F^{-1} \left(\frac{4}{\varepsilon <F^{\sigma 0} >} \right)$  and     $\tilde{E}_{n\bm q\sigma } (\varepsilon _{0n} )=\tilde{\Omega }$. 

With account of differentiation of the inverse function    $F^{-1} \left(x\right)$  we have
\begin{eqnarray} 
\label{energy57} 
\left|(\tilde{\Omega }-\tilde{E}_{n{\bm q}\sigma } (\varepsilon ))^{'} \right|_{\varepsilon =\varepsilon _{0n} } =\left. \frac{4}{\varepsilon ^{2} <F^{\sigma 0} >\left|F^{'} (\varepsilon )\right|} \right|_{\varepsilon =\varepsilon _{0n} }  
\end{eqnarray}
Using (\ref{energy56}) and (\ref{energy57}) one can write the electron  spectral density for \textit{n}-th  band
\[R_{\sigma n} (\Omega ,0)=\]
\[=-\int _{-2}^{2}d\varepsilon  \frac{2\pi }{W} D_{C} \left(\varepsilon \right)\delta (\varepsilon -\varepsilon _{0n} )\frac{F(\tilde{\Omega })\left|F'(\tilde{\Omega })\right|}{F'(\tilde{\Omega })} \varepsilon <F^{\sigma 0} >,\] 
where    $n\omega_0<\Omega<(n+1)\omega_0$  and $\left|F'(\tilde{\Omega })\right|/F'(\tilde{\Omega })=-1$ (see Eq.(\ref{energy54})). We have to combine all \textit{n} bands into one that gives the uniform electron spectral density throughout the whole  frequency interval:
\begin{eqnarray} 
\label{energy58} 
R_{\sigma } (\Omega ,0)=\frac{8\pi }{W} D_{C} \left(\psi(\Omega) \right) 
\end{eqnarray}
where $\psi(\Omega)=4/(\beta W\tilde{G}_{0\sigma }(\Omega )<F^{\sigma 0}>)$.

\begin{figure}[h]
\begin{center}
\includegraphics[width=\columnwidth]{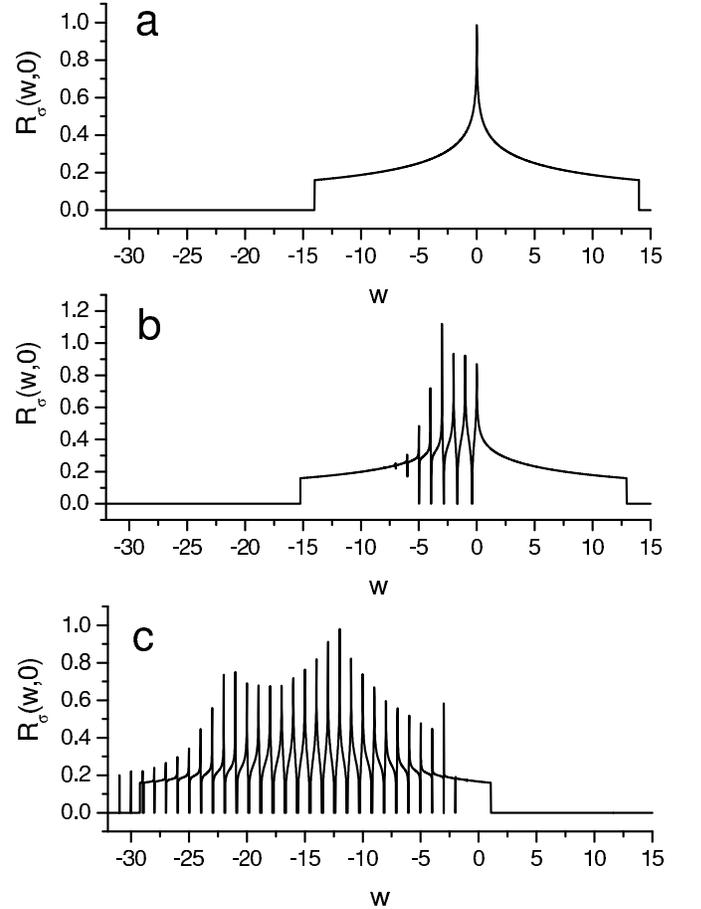}
\end{center}
\caption{The frequency dependences of the electron spectral density $R_{\sigma } (w,0)$ at temperature \textit{T}=0 in units \textit{W } and  g/\textit{W}=0 (a), 0.02 (b) and 0.07(c), where $w=(\Omega -\varepsilon _{\sigma } )/\omega _{0} $.}
\label{fig.5}
\end{figure}

In Fig.~\ref{fig.5} the frequency dependences of the electron spectral density $R_{\sigma } (w,0)$ at temperature \textit{T}=0 in units \textit{W} are presented. In the absence of an electron-phonon interaction it is observed the typical 2d-dimensional spectral density with van Hove singularity. With increase the constant of electron-phonon interction \textit{g} the polaron bands are formed, each of which has a pointed singularity. Also, a whole band is shifted to the left edge and its bandwidth depends on \textit{g} weakly.

\begin{figure}[h]
\begin{center}
\includegraphics[width=\columnwidth]{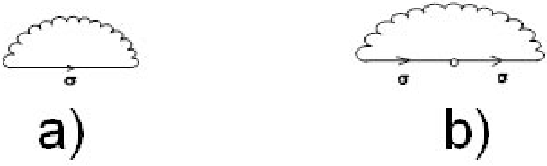}
\end{center}
\caption{ Diagrams for functions $\beta \delta \tilde{\mu }_{\sigma }^{} $ (a)  and $\tilde{\nu }_{\sigma } <F^{\sigma 0} >_{1} $ (b)  from Eq.(\ref{energy46}).}
\label{fig.6}
\end{figure}

Now we will calculate the  functions  $\tilde{\nu }_{\sigma }^{} $ and $\delta \tilde{\mu }_{\sigma }^{} $ in  Eq.(\ref{energy46})   for $\tilde{\mu }$ with account of  the electron-phonon interactions. These   functions correspond to diagrams  \textit{a} and \textit{b} in Fig.~\ref{fig.6}, where the stright and wave  lines  denote  ${\kern 1pt} \tilde{G}_{0\sigma } (i\omega _{n} )$ from (\ref{energy31}) and   effective kinematic interaction $B_{0\sigma } (i\omega _n ,\bm q)$, respectively, and  presented as
\[\beta B_{0\sigma } (i\omega _{n} ,\bm q)=\frac{\beta t(\bm q)}{1-\beta t(\bm q)\tilde{G}_{0\sigma } (i\omega _{n} )<F^{\sigma 0} >} \]
Then one can write
\begin{eqnarray}
\begin{array}{c}\label{energy59} 
\beta \delta \tilde{\mu }_{\sigma }=\frac{1}{N} \sum\limits_{\bm q n}\beta B_{0\sigma } (i\omega _{n} ,\bm q)\tilde{G}_{0\sigma }(i\omega _{n} )\\ \tilde{\nu }_{\sigma } <F^{\sigma 0} >_{1} = \\ 
=\frac{1}{N} \sum\limits_{\bm q nm}\beta B_{0\sigma } (i\omega _{n} ,\bm q)U_{ep} (i\omega _{n} -i\omega _{m} )\cdot\\ 
\cdot(G_{0\sigma } (i\omega _{m} ))^{2} <F^{\sigma 0}>_{1}, 
\end{array}
\end{eqnarray}
where the Fourier component $U_{ep} (i\omega _{n} )$  of   boson unperturbed Green's function $U_{ep} (\tau _{i} -\tau _{j} )$ has a form
\[U_{ep} (i\omega _{n} )=\frac{1}{\beta } \sum _{m=-\infty }^{+\infty }d_{m} \frac{2\sinh \left(\beta m\omega _{0} /2\right)}{i\omega _{n} +m\omega _{0} }  \]
and  $\omega _{n} =4\pi nT$. Also, $\tilde{\nu }_{\sigma } $ in (\ref{energy59}) depends on unperturbed pure electron Green's function (\ref{energy30}), whereas $\delta \tilde{\mu }_{\sigma }^{} $  is determined by $\tilde{G}_{0\sigma } (i\omega _{n} )$ from (\ref{energy31}). The method of inverse function can be used to Eq.(\ref{energy59}) to make a frequency summation. Indeed, in accordance  with formula from \cite{Mahan} one can evaluate the summation by  integration over contour \textit{C} to be a circle of radius R-$>$$\infty$. Then we have
\[\begin{array}{c}\delta \tilde{\mu }_{\sigma }^{} =\frac{-1}{N} \sum\limits_{{\bm q}}\frac{1}{2\pi i} \oint _{{\it C}}\beta B_{0\sigma } (\omega ,\bm q){\kern 1pt} \tilde{G}_{0\sigma } (\omega )f(\omega)d\omega =\\
=\frac{1}{N} \sum\limits_{{\bm q}}\frac{1}{2\pi i} \sum\limits_{k}Res\left[\beta B_{0\sigma } (\omega ,\bm q){\kern 1pt} \tilde{G}_{0\sigma } (\omega )\right]_{\omega _{k} } f(\omega _{k} )\end{array}  ,\]
where $Res\left[\varphi (x)\right]_{\omega _{k} }$is the residue of $\varphi (x)$ in pole $\omega_k$. 

On then applying the similar procedure as for $R_{\sigma } (\Omega ,0)$ above, we  obtain
\begin{eqnarray} 
\label{energy60} 
\delta \tilde{\mu }_{\sigma }^{} =\int _{-\infty }^{+\infty }d\Omega f(\Omega )\frac{4D_{C} \left(\psi(\Omega) \right)}{W<F^{\sigma 0} >^{2} \beta \tilde{G}_{0\sigma } (\Omega )}\ \ \   
\end{eqnarray}

The expressions for $\tilde{\nu }_{\sigma } $ is more complicated:
\begin{eqnarray}
\label{energy61}
\tilde{\nu }_{\sigma } =\tilde{\nu }_{\sigma 1} +\tilde{\nu }_{\sigma 2} +r_{\sigma }\ \ \ \ \ \ \ \ \ \ \ \ \,
\end{eqnarray}
where $\tilde{\nu }_{\sigma i} $  and $r_\sigma$  are  equal
\begin{eqnarray}
\label{energy62}
\begin{array}{c}
{\tilde{\nu }_{\sigma i} =\int _{-\infty }^{+\infty }d\Omega f(\Omega )\frac{4\varphi _{0\sigma }^{i} (\Omega )D_{C} \left(\psi(\Omega) \right)}{W<F^{\sigma 0} >^{2} \left[\tilde{G}_{0\sigma } (\Omega )\right]^{2} }  } \\ {r_{\sigma } =-\frac{1}{<F^{\sigma 0} >} \sum\limits_{m=-\infty }^{+\infty }f(\varepsilon _{\sigma } -m\omega _{0} ) } 
\end{array} 
\end{eqnarray}
Here,
\begin{widetext}
\[\begin{array}{c} {\varphi _{0\sigma }^{1} (\Omega )=-\frac{1}{\beta } \sum\limits_{m=-\infty }^{+\infty }\frac{2d_{m} f(\varepsilon _{\sigma } )(1-f(\varepsilon _{\sigma } ))\sinh \left(\beta m\omega _{0} /2\right)}{\Omega +m\omega _{0} }  \approx } \\ {\quad \quad \approx -\frac{1}{\beta \Omega } f(\varepsilon _{\sigma } )(1-f(\varepsilon _{\sigma } ))\left\{M(1,1+\Omega /\omega _{0} ,-\lambda ^{2} )-M(1,1-\Omega /\omega _{0} ,-\lambda ^{2} )\right\}} \\ {\varphi _{0\sigma }^{2} (\Omega )=-\frac{1}{\beta ^{2} } \frac{\partial (\beta \tilde{U}_{0\sigma } (\Omega ))}{\partial \Omega } =-\frac{1}{\beta ^{2} \omega _{0} } \frac{\partial (\beta \tilde{U}_{0\sigma } (w))}{\partial w} } \end{array}\]
\end{widetext}

The function (\ref{energy61}) it seems to be divergent since we have in (\ref{energy62}) the sum of  Fermi functions. However, the detailed analysis shows that pointed  divergence is absent. To prove this statement we divide the area of integration on sections from  one  singularity \textit{m}  to other \textit{m}+1 and then the sum over all intervals is written as
\begin{eqnarray}
\label{energy63}
\tilde{\nu }_{\sigma 2} +r_{\sigma } =\frac{1}{<F^{\sigma 0} >} \sum\limits_{m=-\infty }^{+\infty }\left\{R_{m} -f(\varepsilon _{\sigma } -m\omega _{0} )\right\},
\end{eqnarray}
where the integral on \textit{m}-th segment has a form
\begin{eqnarray} 
\label{energy64}
R_{m} =\int _{m}^{m+1}dwf(w\omega _{0} +\varepsilon _{\sigma } )\frac{\partial }{\partial w}  \left(\psi(w) \right)D_{C} \left(\psi(w) \right)\ \ 
\end{eqnarray}

Let $w_{lm}$ and $w_{rm}$ be the left and right edges of the segment with \textit{m}-th singularity, respectively. They are determined by equations:
\[\frac{4}{W\beta \tilde{G}_{0\sigma } (w_{im} )<F^{\sigma 0} >} =\pm 2, \]
where upper and lower signs  correspond to $w_{lm} $ and $w_{rm} $, respectively. On now carrying out the partial integration  in (\ref{energy64}) it is easy to obtain for $R_{m} $: 
\begin{eqnarray} 
\label{energy65} 
R_{m} =\frac{1}{2} \left[f(w_{lm} \omega _{0} +\varepsilon _{\sigma } )+f(w_{rm} \omega _{0}+\varepsilon _{\sigma } )\right]-\nonumber\\
-\int _{m}^{w_{lm} }dw\frac{df(w\omega _{0} +\varepsilon _{\sigma } )}{dw}  S\left(D_{C} \left(\psi(w) \right) \right)-\nonumber\\  -\int _{w_{lm} }^{m+1}dw\frac{df(w\omega _{0} +\varepsilon _{\sigma } )}{dw}  S\left(D_{C} \left(\psi(w) \right) \right)  
\end{eqnarray}
Here, \textit{S(x)= I(x)-0.5} (see Eq.(\ref{energy49})). Apparently, when\textit{ m-}$>$$\pm$$\infty$   the frequency  $w_{lm} \to w_{rm} $ and Fermi functions from  Eq.(\ref{energy65}) will compensate ones from Eq.(\ref{energy63}).  It reflects the existence at high frequencies the localized phonon modes only.   One can point out that the number of polaron bands is not exceed  35-40 (see Fig.~\ref{fig.5}, c)  for considered temperature area and \textit{g/W}$<$0.085. Finally, the Eqs.(\ref{energy60}), (\ref{energy61}) and (\ref{energy65}) allow to find the numerical solution of  a set of equations (\ref{energy44}) and (\ref{energy46}) without any difficulties.

\section{RESULTS OF THE NUMERICAL CALCULATIONS}

To calculate the temperaure of superconducting state (\ref{energy44}) we take the value of hopping integral $t\sim $0.5 eV  and exchange parameter $J\sim $0.023 eV \cite{Kulic,Cyrot}. It corresponds the cuprate YBa$_2$Cu$_3$O$_6$ with Neel temperature $\ \  T_N$=420 K. The parameter $J$  is determined with account of  $\sim 30\%$ contribution  of the spin fluctuations. Then we have the bandwidth $W$=4 eV. Also, we take the frequency $\omega$=75 meV to be typical for cuprates in Einstein model for phonons \cite{Verga}. Finally, we have $\omega_0/W$=0.01875 and $J/W$=0.058.

Let us consider the solutions of Eq.(\ref{energy46}) at temperaure $\textit{T}$=0.  At $T\sim$0 one can rewrite Eq.(\ref{energy46}) as
\begin{eqnarray} 
\label{energy66} 
\tilde{\mu }+\delta {\kern 1pt} \tilde{\mu }_{\sigma } =T\ln \frac{n/2-\tilde{\nu }_{\sigma } }{1/2-n/2+\tilde{\nu }_{\sigma } /2}  
\end{eqnarray}
It follows from (\ref{energy66}) that at \textit{T}=0  for PM-2 phase, when $\tilde{\mu }+\delta {\kern 1pt} \tilde{\mu }_{\sigma } >0$, the chemical potential obeys the equation  $1/2-n/2+\tilde{\nu }_{\sigma } /2=0$. Surprisingly, but there is a situation, when $\tilde{\mu }+\delta {\kern 1pt} \tilde{\mu }_{\sigma } =0$ and $1/2-n/2+\tilde{\nu }_{\sigma } /2\ne 0$, $n/2-\tilde{\nu }_{\sigma } \ne 0$.

\begin{figure}[h]
\begin{center}
\includegraphics[width=\columnwidth]{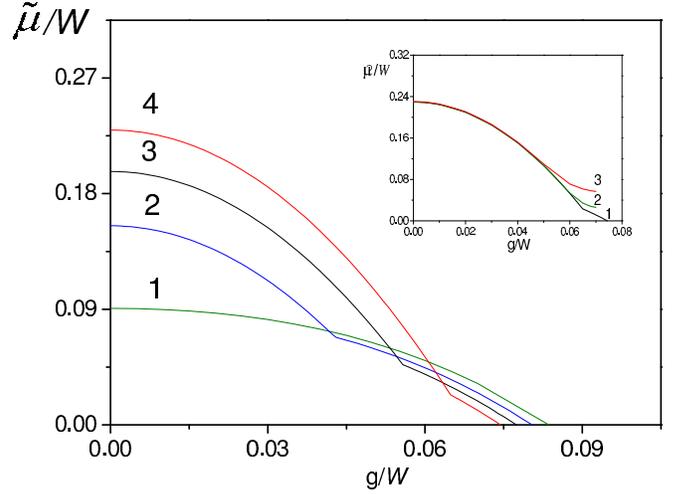}
\end{center}
\caption{ The chemical potential versus costant of the electron-phonon interaction \textit{g} at electron concentrations \textit{n}=0.571, 0.7, 0.8 and 0.9 (curves 1-4, respectively) and temperature \textit{T}=0. In inset: $\tilde{\mu }$ as function of \textit{ g} at \textit{n}=0.9 and  temperatures \textit{T/W}= 0, 0.005 and  0.0136 (curves 1-3, respectively).}
\label{fig.7}
\end{figure}

In Fig.~\ref{fig.7} the curves of  $\tilde{\mu }$ as function of $g$ at different electron concentrations are presented. On curves 2-4 at $g=g_{cr.}$ one can observe the kinks. When $g>g_{cr.}$ the chemical potential obeys the equation $\tilde{\mu }+\delta {\kern 1pt} \tilde{\mu }_{\sigma } =0$ and  at $g<g_{cr}$. for $\tilde{\mu }$ we have $1/2-n/2+\tilde{\nu }_{\sigma } /2=0$. At critical concentration  $n=n_{cr.}$   of  transition in PM-1 phase we have $g_{cr}$.=0. In inset the influence of temperature is shown.

\begin{figure}[h]
\begin{center}
\includegraphics[width=\columnwidth]{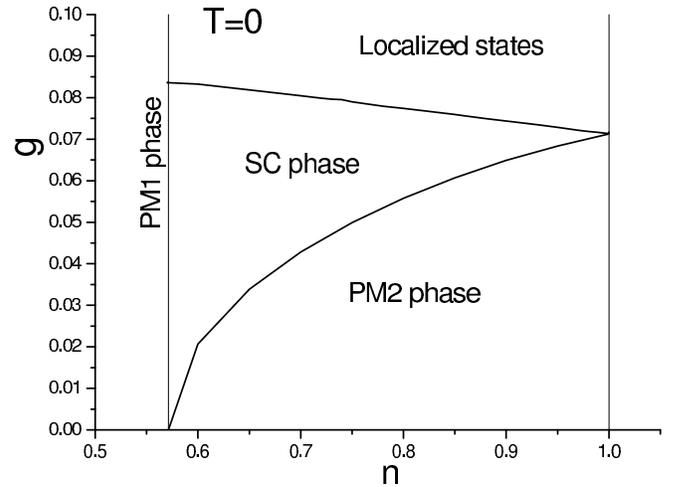}
\end{center}
\caption{ Phase diagram  of the strongly correlated electron system with electron-phonon coupling.}
\label{fig.8}
\end{figure}

In Fig.~\ref{fig.8} the phase diagram in coordinate \textit{g-n} at \textit{T}=0 is presented. The upper and  lower  parts  of  boundary  of superconducting state (SC) is detrmined by $g_{cr}$.  as function  of  \textit{n}  and  value  of  \textit{g } at which $\tilde{\mu }=0$ ( see   Fig.~\ref{fig.7}). In SC phase the contribution of the polaron excitation in the narrowing of electron band is essential. The chemical potential is decreased up to zero. Thus, the condition (\ref{energy45})  for  appearance of  SC state can be realized.  In the area of localized state we can not obtain the solution for $\tilde{\mu }$  corresponding to hole doped electron system (PM-2 phase).  

\begin{figure}[h]
\begin{center}
\includegraphics[width=\columnwidth]{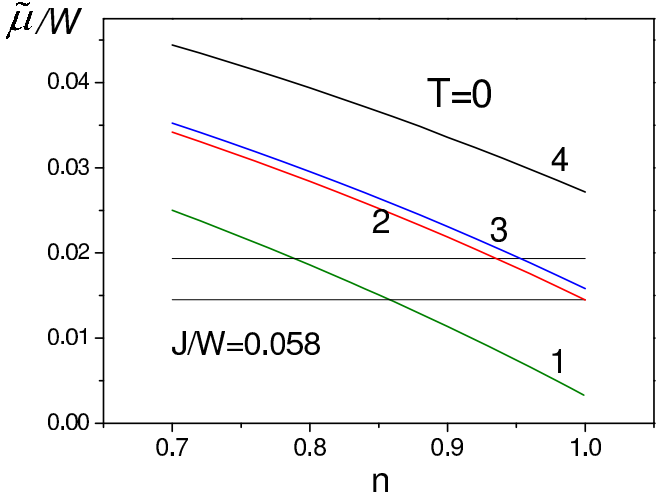}
\end{center}
\caption{ The concentration dependences of the chemical potential  at \textit{T}=0 and  parameters of  electron-phonon coupling  \textit{g/W}= 0.07, 0.06555, 0.065 and 0.06 (curves 1-4, respectively).}
\label{fig.9}
\end{figure}

In Fig.~\ref{fig.9} the concentration dependences of the chemical potential  for different parametrs of  electron-phonon coupling are presented at \textit{T}=0. The srtright lines shows  a region of acceptability of   $\tilde{\mu }$ at different values of  \textit{g }and \textit{n}. One can see that at small values of \textit{g}  the bandwidth is unsufficiently narrow. At large values of  \textit{g}  the pointed region is shifted to high hole concentration,  where the effects of kinematic interactions are strong.

\begin{figure}[h]
\includegraphics[width=\columnwidth]{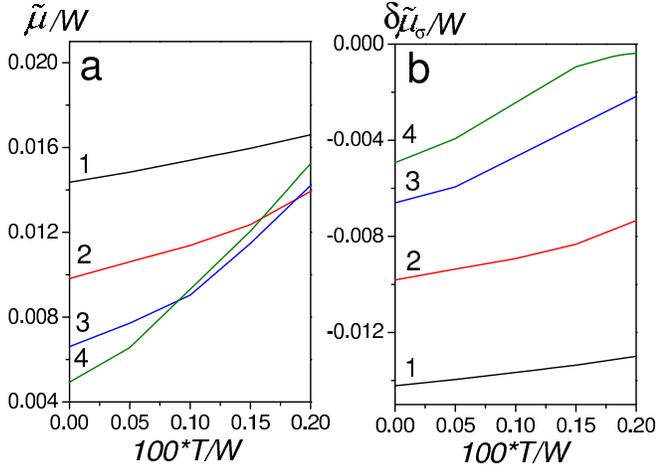}
\caption{The temperature dependences of $\tilde{\mu }$ (a) and $\delta \tilde{\mu }_{\sigma } $ (b)  at \textit{g/W}=0.07 and \textit{n}=0.86, 0.92, 0.96 and 0.98 (curves 1-4, respectively).}
\label{fig.10}
\end{figure}

Now we will calculate the temperature of  superconducting state $T_{SC}$. In  accordance with Eq. (\ref{energy44}) it is necessary to find $\tilde{\mu }$ and $\delta \, \tilde{\mu }_{\sigma } $ as  functions of  \textit{T} and electron concentration. In Fig.~\ref{fig.10} the temperature dependences of $\tilde{\mu }$ and $\delta \, \tilde{\mu }_{\sigma } $ at \textit{g/W}=0.07 and different electron concentration were obtained   by solving the Eq.(\ref{energy46}). From Fig.~\ref{fig.10},\textit{a}  it easy to see  that chemical potential in a definite temperature and concentration area meets the condition (\ref{energy45}) and as a result one can solve the Eq.(\ref{energy44}) for $T_{SC}$.  

\begin{figure}[h]
\begin{center}
\includegraphics[width=\columnwidth]{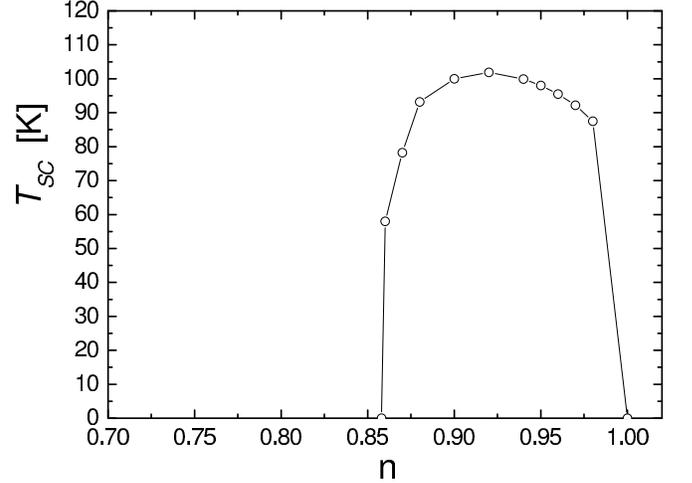}
\end{center}
\caption{ The temperature of  superconducting state $T_{SC}$ versus electron concentration \textit{n}  with strength of electron- phonon coupling \textit{g/W}=0.07  and bandwidth \textit{W}=4 eV.}
\label{fig.11}
\end{figure}

In  Fig.~\ref{fig.11} the concentration dependence of  the temperature  of superconducting state is presented  for \textit{g/W}=0.07 and \textit{W}=4eV. We observe the high-temperature superconductivity with left and right edge of  superconducting phase on axis \textit{n}.
The area  between stright lines determines  $\tilde{\mu }$ in accordance with requirement (\ref{energy45}) for SC phase. Indeed, the left edge of SC phase $n_{LSC}$=0.858 when $T_{SC}(n_{LSC})$=0 is determined  from equation $J/4+\delta \, \tilde{\mu }_{\sigma } =0$ (see Eq.\ref{energy60}), i.e.
\[\frac{4(\omega _{0} /W)^{2} }{(1-n/2)^{2} } \int _{-\infty }^{\tilde{\mu }/\omega _{0} }dw\cdot w\frac{4D_{C} \left(\frac{4w\omega _{0} /W}{(1-n/2)M(1,1+w,-\lambda ^{2} )} \right)}{M(1,1+w,-\lambda ^{2} )}=\]\[=-J/(4W)\] 

The right edge $n$=1 when  $T_{SC}$=0 follows from condition $\delta \, \tilde{\mu }_{\sigma } /W\to 0$ at $n\to 1$(see Fig.~\ref{fig.10}, b). It reflects the weakening of  effective kinematic field at half filling of the  band near $T_{SC}$. Since $\tilde{\mu }\to J/4$ then $\ln \frac{J-2\tilde{\mu }}{4\tilde{\mu }-J} \to +\infty $ and $T_{SC}$-$>$0. The maximum of  $T_{SC}$ from Fig.~\ref{fig.11}  is in a good agreement with experimental value $T_{SC}\sim$100K for cuprate YBa$_2$Cu$_3$O$_7$ \cite{Kulic}.

It is interesting to find the gap  as a functions of  \textit{n }at\textit{ T}=0 and  its temperature dependence for fixed \textit{n}. The gap $\Delta$ is determined from Eq.(\ref{energy41}) by expansion $\left|\tilde{{\it E}}_{{\it q}} \right|$ into a series in $\Delta /\tilde{\mu }$.  It is easy to find  that
\[\Delta _{k} =<F^{\sigma 0} >_{1} \Delta _{k} \frac{J}{2\tilde{\mu }} \left\{1-\frac{9}{8} \frac{\Delta _{}^{2} }{\tilde{\mu }^{2} } +\frac{75}{32} \frac{\Delta _{}^{4} }{\tilde{\mu }^{4} } -...\right\},\]
where $<F^{\sigma 0} >_{1} $ is determined by (\ref{energy42}) with $\varepsilon _{\sigma } =-\mu _{g} $. At $T$=0 we have $<F^{\sigma 0} >_{1} $=1/2  and  it gives  the algebraic equations for  $\Delta /\tilde{\mu }$. For example, at $n$=0.9 and $g/W$=0.07 from Fig.~\ref{fig.7} (curve 4) we have $\tilde{\mu }/W$=0.0114 and $\Delta/W $=0.0055. With account of $T_{SC}/W$=0.00215 at $n$=0.9 we obtain $2\Delta/T_{SC}$=5.15. By similar manner the gap $\Delta $ versus $T$  was calculated.

\begin{figure}[h]
\begin{center}
\includegraphics[width=\columnwidth]{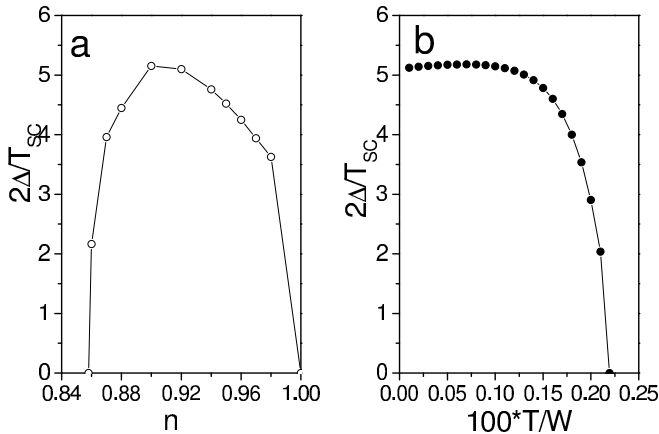}
\end{center}
\caption{Relationship between gap and $T_{SC}$ versus electron concentration $n$  at zeroth temperature (a). Temperature dependence of  $2\Delta /T_{SC} $  at $n$=0.92, $g/W$=0.07, $J/W$=0.058 and $T_{SC}/W$=0.0022 (b).}
\label{fig.12}
\end{figure}

In Fig.~\ref{fig.12}  the concentration (a) and temperature (b)  dependencies   of the relationship $2\Delta /T_{SC} $ are presented. An optimal doping  gives the maximal value of  
$2\Delta /T_{SC}\approx$ 5 that also corresponds to experimental results for cuprates \cite{Kulic}.
\section{CONCLUSIONS}     
In the present work, an influence of the electron-phonon interaction on normal and superconducting  properties  of  the  strongly correlated electrons  has been studied. The strong kinematic interactions in a doped system were shown to destroy the singlet pairs of electrons formed by indirect exchange. The inclusion of the sufficiently strong electron-phonon coupling stabilizes these pairs by virtue of  correlation band  narrowing manifested in a drastic decrease of the chemical potential. It was built the phase diagram of ground states in coordinates the constant of electron-phonon coupling  \textit{g}  and electron concentration \textit{n}. At optimal value of \textit{g} and \textit{n} for cuprate YBa$_2$Cu$_3$O$_7$ the calculated critical temperature of superconductivity $T_{SC}$ and  2$\Delta/T_{SC}$ are very close  to experimentally observed. This value $g/W$=0.07 corresponds to energy of the Holstein's polaron $E_p=g^2/\omega_0\approx$ 1.05 eV and its radius $R_p=a(W/E_p)\approx4a$. A new method of the frequency summation with infinite number of poles of unperturbed Green's function  was suggested. The exact analytic expressions  of  diagrams in the first nonvanishing approximation of perturbation theory  were obtained to find the contribution of polarons in the chemical potential.


\begin{references}

\bibitem{Shin}
 Shin-ichi Uchida,
 Jap.\ Journ.\ App.\ Phys.\ \textbf{51}, 010002 (2012).

\bibitem{Gros01}
 C. Gros, R. Joynt, and T.M. Rice,
 Phys.\ Rev.\ B\ \textbf{36}, 381 (1987).

\bibitem{Baskar}
G. Baskaran, Z. Zou and P.W. Anderson,
Sol.\ St.\ Comm.\ \textbf{63}, 973 (1987).
%
\bibitem{Kulic}
M.L. Kuli\^c,%
Physics Reports \textbf{338}, 1 (2000).
\bibitem{Ciuchi}
 S. Ciuchi,  F. Pascuale, S. Fratini, D. Feinberg,
 Phys.\ Rev.\ B\ \textbf{56}, 4494 (1997).
\bibitem{Alexandr}
 A. Alexandrov and J. Ranninger,
 Phys.\ Rev.\ B\ \textbf{24}, 1164 (1981).
\bibitem{Lang}
 I.G. Lang and  Yu.A. Firsov,
 Zh.\ Exp.\ Theor.\ Fiz.\ \textbf{43}, 1843 (1962)(Sov.\ Phys. \ JETP \textbf{16}, 1301 (1963)).
\bibitem{Izyum01}
 Yu.A. Izyumov, B.M. Letfulov, E.V. Shipitsyn, M. Bartkowiak, K.A. Chao,
 Phys.\ Rev.\ B\ \textbf{46}, 15697 (1992).
\bibitem{Izyum02}
 Yu.A. Izyumov, M.I. Katsnelson, Yu.N. Skryabin, 
\textit{Magnetism of the collectivized electrons}, (Fizmatgiz, Moscow, 1994), p.366.
\bibitem{Zubov01}
 E.E. Zubov, V.P. Dyakonov, H. Szymczak,
 J.\ Phys.:\ Condens.\ Matter.\ \textbf{18}, 6699 (2006).
\bibitem{Mahan}
 G.D. Mahan, 
\textit{Many Particle Physics}, (Plenum, New York, 1990), p.793.
\bibitem{Zubov02}
 E.E. Zubov,  
 Physics-\ Solid\ State\ \textbf{51}, 106 (2009).
\bibitem{Myron}
 S.F. Myronova, E.E.  Zubov,
 J.\ Magn.\ Magn.\ Mater.\ \textbf{316}, e274 (2007).
\bibitem{Zubov03}
 E.E. Zubov, 
 Theoretical\ and Mathematical\ Physics\ \textbf{105}, 1442 (1995).
\bibitem{Brinkman}
 W.F. Brinkman and  T.M.Rice,
 Phys.\ Rev.\ B\ \textbf{2}, 1324 (1970).
\bibitem{Edwards}
 D.M. Edwards,
 Adv.\ in Phys.\ \textbf{51}, 1259 (2002).
\bibitem{Abram}
 M. Abramowitz and C.A. Stegun, (Eds.), 
\textit{Handbook of Mathematical Functions with Formulas, Graphs, and Mathematical Tables 
}, (Dover, New York, 1972), p.1046.
\bibitem{Cyrot}
 M. Cyrot,
 Sol.\ St.\ Comm.\ \textbf{62}, 821 (1987).
\bibitem{Verga}
  S. Verga, A. Knigavko, and F. Marsiglio,
 Phys.\ Rev.\ B\ \textbf{67}, 054503 (2003).
\end{references}
\end{document}